\documentclass[aps, prb, twocolumn,amsmath,amstex,amssymb,mathfonts,superscriptaddress,longbibliography]{revtex4-2}

\usepackage{subfigure}
\usepackage{amssymb}
\usepackage{amsmath}
\usepackage{graphicx}
\usepackage{dcolumn}

\usepackage{physics}
\usepackage{bm}
\usepackage[colorlinks, linkcolor=cyan, anchorcolor=blue, citecolor=green]{hyperref} 

\usepackage{stackengine}
\newcommand{\doublesubring}[1]{\ensurestackMath{\stackon[-12pt]{#1}{\mkern2mu\circ\circ}}}
\renewcommand{\textsubring}[1]{\ensurestackMath{\stackon[-12pt]{#1}{\mkern2mu\circ}}}
\newcommand{\blankchar}{\color{white}{1}}

\usepackage{color}

\usepackage{amsmath}
\usepackage{amssymb}
\usepackage{amsthm}
\usepackage{amsfonts}
\usepackage{multirow}
\usepackage{makecell}
\usepackage{float}
\usepackage{comment}
\usepackage{enumitem}
\usepackage{verbatim}
\usepackage{mathtools}

\begin{document} 

\title{Dynamics of Anyon Clusters in Fractional Quantum Hall Fluids}

\author{Qianhui Xu}
\affiliation{Division of Physics and Applied Physics, Nanyang Technological University, Singapore 637371}

\author{Guangyue Ji}
\affiliation{Department of Physics, Temple University, Philadelphia, Pennsylvania, 19122, USA}

\author{Yuzhu Wang}
\author{Ha Quang Trung}
\author{Bo Yang$^*$}
\affiliation{Division of Physics and Applied Physics, Nanyang Technological University, Singapore 637371}

\date{\today}
\begin{abstract}
In fractional quantum Hall fluids, the quasiparticle excitations are anyons with fractional charges and statistics. Effective interactions among the anyons can be induced by either model or realistic electron-electron (e-e) interactions. Without losing the generality, we investigate such phenomena for the Laughlin $1/3$ and Moore-Read (MR) non-Abelian phases. Anyons display rich internal dynamics in both cases that will lead to interesting experimental consequences. In particular, bound states of two Laughlin anyons are preferred under short-range e-e interactions, leading to $2e/3$ -- instead of $e/3$ -- effective charge carriers at low temperatures, which have been seen in several experiments. MR phases host two topologically distinct fusion channels: “1” and “$\psi$”. The different effective interactions of $e/4$ anyons in the two sectors suggest the vanishing of the degeneracy of fusion channels when the e-e interaction is no longer its model Hamiltonian, in which case different bound states could also appear. This indicates the possibility of energetically manipulating the two types of anyons by tuning the bare e-e interactions. We point out that the results of recently developed high-resolution STM measurements will be affected by the effective anyon interactions, where anyons are clustered together after the tunneling of electrons. The low-lying parts of the local density of states affected by various anyon clusters are simulated for both Abelian and non-Abelian systems with (screened) Coulomb interactions.
\end{abstract}

\maketitle 

\section{Introduction}
Anyons with fractional charges and fractional statistics are particles only existing in two-dimensional manifolds because of the nontrivial braid group in 2D \cite{Wilczek82QMforAnyons, wilczek82magnetic, wilczek1985statistics, MooreRead1991,wu1991braid, Douglas2023anyons}. The fractional quantum Hall (FQH) states, realized in a two-dimensional electron gas (2DEG) with strong electronic interaction subjected to a perpendicular strong magnetic field at low temperature, provide an ideal platform to study the exotic properties of anyons \cite{Wilczek1984statistics&FQH,halperin84Statistics, sharma2020bardeen, scarola2000cooper}. Quasiholes that can be regarded as anyons in the incompressible FQH fluids are low-lying collective excitations of electrons created by adiabatically inserting fluxes into the FQH ground states with distinct topological orders \cite{Laughlin1983, wen1995TOandFQH}. The simplest example is the Laughlin $1/m$ anyons with fractional statistics $\nu \pi$ and fractional charge $\nu e$, created by inserting one flux into the Laughlin ground state at filling factor $\nu=1/m$ \cite{Kjonsberg1997}. More intriguing cases are the non-Abelian phases in which several anyons with non-Abelian statistics are created by one flux insertion. The great interest in non-Abelian anyons is rooted in the possibility of their application to fault-tolerant topological quantum computing \cite{Kitaev2003fault,nayak2008NA, Bernard2018TQC}. 

Much effort has been made to understand the properties of emergent quasi-particles in FQH fluids, such as their fractional statistics, the relationship between them in different FQH phases, the size of anyons and their structures, etc \cite{S.Simon2009MR,Bernevig2014braiding, Trung2023spin-statistics, Trung2021dynamics, Papic2014size, guangyue2024oscillation}. One modern perspective is that anyons are ``elementary particles" within the respective conformal Hilbert spaces (CHSs) defined by model Hamiltonians, or more generally, by local exclusion conditions (LEC) \cite{MooreRead1991,girvin1987quantum,simon2007projection,hansson2017revQHCFT, boyang2019emergent, boyang2021ClassicalConstraints}. These CHSs with emergent conformal symmetry are spanned by the ground state and all quasihole states \cite{BY2022anyons,yuzhu2023geometric}. If only the model Hamiltonian defining the CHS exists as the electronic interaction term, then the anyons are massless (i.e., it costs no energy to create a single anyon) and free (i.e., the anyons are non-interacting) particles. With more realistic electron interactions, they will gain self-energy and start to interact \cite{anyon2001interaction}. Previous studies manifest the hierarchical structure of CHSs, implying the nontrivial internal structures of anyons as the elementary particles in the corresponding CHSs \cite{BY2022anyons, Read2009conformal}. These intricate properties and the nature of CHSs serve as the foundation for revealing the exotic properties of different topological FQH phases and their relationship \cite{BY2022anyons, Trung2021dynamics,yuzhu2023geometric, yuzhu2022analytic, LingjieDu2024graviton}. 

Experimental progress, such as shot noise, quantum point contact, and anyon interferometry, provides direct detection for both Abelian and non-Abelian natures of certain anyons \cite{Feve2020exp-Collisions,sim2022exp-NAcollider,feve2023exp-comparing,lee2023exp-partitioning, Pierre2023exp-statistics, stern2018thermalHall, thomas2025anyon}. While these experimental methods often require a dilute anyon gas for braiding and quantum tunneling, meaning a large separation between anyons, other phases such as Wigner crystal, stripes, and bubbles will compete with FQH phases when the anyon's density is high \cite{MacDonald1984solid, MacDonald1985solid,2016solid,2018bubble,yuzhu2023dr}. This leads to the rich phase diagrams of 2DEG and makes it interesting to understand the dynamics of a cluster of anyons when they are close and their interaction becomes important. Recent advancements in atomic-scale spectroscopy within ultra-clean, diverse 2D systems have enabled the use of local scanning tunneling microscopy (STM) to measure tunneling processes of a cluster of anyons, probe anyon clusters' dynamics, and provide valuable insights into their energies and fusion rules \cite{exp_STM_Papic2018, exp_feldman2021, exp_STM_coissard2022imaging, exp_STM_highReso}. In such experiments, the proximity of anyons amplifies the role of effective interactions that strongly influence their behavior \cite{Jain2024STM}.

In this work, we focus on the effective microscopic interaction between anyons in the Laughlin and Moore-Read (MR) phases, derived from both the model and the realistic bare interactions between electrons. Such effective interactions are highly nontrivial, reflecting information about the particle statistics\cite{wilczek1985statistics, Wilczek1984statistics&FQH} and the spatial extension of particle sizes (i.e., anyons are not point particles)\cite{boyang_anyon, guangyue2024oscillation}. This leads to the formation of few-anyon clusters and are potentially important for many-anyon quantum fluids with both universal and non-universal dynamics. In particular, we show that two Laughlin 1/3 anyons form a bound “molecule” with short-range electron-electron (e-e) interactions, providing a possible explanation for the anomalous results in quantum point contact experiments of Laughlin phases \cite{efcharge-PRL2009shot,efcharge-PRB2010,efcharge-PRL2011,efcharge-NP2017}. Moreover, different dynamics of non-Abelian $\psi$-type and Abelian $1$-type anyons in the Moore-Read phase with the same e-e interactions break the ground state degeneracy, implying that the Majorana fermion is no longer massless in the presence of the realistic interactions. Here the term "dynamics" is used in the general sense for the parametric evolution of quasihole clusters' energies, which will manifest themselves in various experiments where dynamical properties of the quantum Hall states are measured. For both Abelian and non-Abelian anyons, we demonstrate that the local density of states (LDOS) detected by high-resolution STM depends on the interactions within the anyon clusters created by adding or removing a single electron. This makes STM a powerful tool for probing the effective anyon interactions in topological FQH phases.

This paper is structured as follows. Section \uppercase\expandafter{\romannumeral2} first briefly reviews the formalism of the microscopic theory for FQHE on spherical geometry, including Haldane pseudopotentials and the Jack polynomial method. Then the dynamics of two- and three-anyon clusters of the Laughlin 1/3 kind induced by both model and realistic electron-electron interactions is studied. Similarly, in section \uppercase\expandafter{\romannumeral3}, the dynamics of two- and four-anyon clusters for the MR kind is investigated, and the distinct interacting behaviors of the two types of MR anyons in systems with odd and even electron numbers under the same electron-electron interaction are highlighted. Finally, in \uppercase\expandafter{\romannumeral4}, we explore the potential experimental signatures of the anyon dynamics, particularly in high-resolution tunneling measurements.
  
\section{Interactions within a cluster of Laughlin quasiholes}
\subsection{Spherical Geometry and Jack Polynomial Formalism}
Numerics in this paper are conducted on a sphere enclosing a Dirac magnetic monopole at the center. If the monopole has total flux $2S$ in units of the magnetic flux quanta $\Phi = h/e$, where $2S$ must be an integer \cite{dirac1931}, it will generate a uniform magnetic field normal to the surface of the sphere whose radius is $R = l_B \sqrt{S}$. $l_B = \sqrt{\hbar/eB}$ is the magnetic length, serving as the fundamental length scale in quantum Hall (QH) systems under a perpendicular magnetic field whose magnitude is $B$. Electrons on the sphere have the angular momentum of $l = |S| + n$, $n=0,1,2,...$ being the LL index. Single-particle orbitals in one LL are the eigenstates of the angular momentum operators $\hat{L}^2$ and its azimuthal part $\hat{L}_{z}$, indexed by $s=-l, -l+1, ..., l$. So a single-particle quantum state can be labeled by $\ket{l,s}$. For a many-body system, the total angular momentum operator and its components are defined as $\hat{L}^2 = \hat{L}^2_x + \hat{L}^2_y + \hat{L}^2_z$, $\hat{L}_\alpha = \sum_i\hat{L}_{\alpha, i}$ where $\alpha=x,y,z$ and $i$ is the particle index.

Assuming the electrons are fully spin-polarized, an isotropic two-body interaction in the $n_{th}$ LL, neglecting LL mixing, can be written as
\begin{equation}
    \label{eq_int}
    \hat{H}_{\text{int}} = \frac{1}{2} \sum_{s_1+s_2 \atop =s_3 +s_4} \bra{s_1,s_2} \hat{V}^{(n)} \ket{s_3, s_4} \hat{c}_{s_1}^\dagger \hat{c}_{s_2}^\dagger \hat{c}_{s_3} \hat{c}_{s_4}
\end{equation}
where $\hat{c}_{s_r}^\dagger$ ($\hat{c}_{s_r}$) creates (annihilates) an electron in the single-particle orbital labeled by $s_r$, with $r$ being the particle index. $\hat{V}^{(n)}$ denotes the interaction projected to the $n_{th}$ LL. It can be expanded in the coupled basis $\ket{L, M}$, where $L=0,1,...,2l$ is the total angular momentum of the electron pair and $M$ is its azimuthal part. Introducing the relative angular momentum of the electron pair $m \equiv 2l-L$, it isolates the relative motion of the pair from its center-of-mass motion, shown in the two-particle wavefunction in spinor coordinates $(u,v)$ on the sphere:
\begin{equation}
    \psi_{(\alpha,\beta)}^{l,m} = (u_1v_2-u_2v_1)^{m} \prod_{i=1,2}\left( \bar{\alpha}u_i + \bar{\beta}v_i \right)^{2l-m}
    \label{eq_wf}
\end{equation}
where $(\alpha, \beta)$ specifies the angular position of the electron pair's center of mass and satisfies $|\alpha|^2 + |\beta|^2 = 1$ \cite{haldane1983hierarchy, 2011_LLonSphere}. $|u_1v_2-u_2v_1| = \sin{(\Omega_{1,2}}/2)$, $\Omega_{1,2}$ is the geodesic angle between the two particles on the sphere. The chord distance between the electrons is $2R |u_1v_2-u_2v_1|$, while the geodesic distance is $ 2R \arcsin{|u_1v_2-u_2v_1|}$. From Eq. (\ref{eq_wf}), $m$ is restricted to be odd (even) for fermionic (bosonic) FQH systems due to the corresponding exchange symmetry \cite{hsiao2020landau, zhu2015fractional}. Smaller $m$ corresponds to configurations with smaller mean separation, while larger $m$ significantly enhances the amplitude for configurations with larger separation, thus characterizing the effective range of the interaction.

The projected interaction can then be decomposed as:
\begin{equation}
    \hat{V}^{(n)} = \sum_{m=0}^{2l} c^{(n)}_m \hat{V}_m
\end{equation}
$c^{(n)}_m = \bra{2l-m, M} \hat{V}^{(n)} \ket{2l-m, M}$ is the pseudopotential (PP) coefficient, representing the interaction energy of two electrons with relative angular momentum $m$ \cite{haldane1983hierarchy, fano1986configuration}.
\begin{equation}
    \hat{V}_m = \sum_{M=-L}^{L}\ket{L, M} \bra{L, M}
\end{equation}
is the PP operator, which projects any two-particle state onto the subspace of relative angular momentum $m$. Unless otherwise stated, all analyses in this paper are performed in the lowest LL (LLL), corresponding to the strong magnetic field limit, and the LL index $n$ is omitted hereafter. PPs provide both an intuitive physical picture and a sparse representation of the interaction, thereby simplifying numerical calculations. More importantly, they define model Hamiltonians whose eigenstates are the model wavefunctions of certain FQH phases. For instance, the eigenstates of the model Hamiltonian $\sum_{k}^p \hat{V}_{2k-1}$, $k=1,2,...$, are the fermionic Laughlin model wavefunctions, describing both the ground states and anyonic quasihole states at the filling factor $\nu = 1/(2p+1)$. These states constitute the null space of the model Hamiltonian $\sum_{k}^p \hat{V}_{2k-1}$ and span the corresponding conformal Hilbert space (CHS). Understanding the effects of individual PPs can also help us predict the influence of realistic interactions that are linear combinations of PPs.

Jack polynomial formalism expresses the many-body wavefunctions of FQH states by root configurations with different admissible rules\cite{bernevig2008Jack, bernevig2008generalizedJack}. Root configurations consist of “1” or “0”, each representing a single-particle orbital on the sphere with azimuthal angular momentum $s = -l, -l+1, ..., l$ from the leftmost (north pole of the sphere) to the rightmost (south pole of the sphere). “1” indicates an orbital occupied by one electron, whereas “0” represents an unoccupied orbital. Laughlin 1/3 states satisfy (1,3) admissibility rule, meaning that there is no more than 1 electron in 3 consecutive orbitals. The ground state for Laughlin 1/3 conformal Hilbert space $\mathcal{H_L}$ is the highest density state\cite{boyang-2021-statistical}:
\begin{equation}
    1001001...001001
\end{equation}
where the number of orbitals $N_o = 3 N_e -2$, $N_e$ being the electron number. Electrons in the ground state symmetrically occupy the orbitals, $L = L_z = 0$. All the other quasihole excitation states in $\mathcal{H_L}$ can be obtained by inserting fluxes ($0s$) into the ground state.

In Abelian phases such as the Laughlin states, one anyon is created by each flux insertion, namely the number of quasiholes $N_{qh}$ equals the number of extra fluxes. The new orbital indices after the insertion will be $s' = -(l+N_{qh}/2), -(l+N_{qh}/2)+1, ..., l+N_{qh}/2$. When all fluxes are inserted at the north pole, each electron's spin $s$ will
increase $N_{qh}/2$, yielding the state with total angular momentum $L=L_{\text{max}} = N_{qh} N_e/2$, such as Eq.(\ref{eq_l_NN}) in which two more fluxes are added. Empty circles below the digits denote the positions of anyons. We define the relative angular momentum of an anyon cluster in a state with total electron angular momentum $L$ as: $\Delta M = L_{max}-L$. Pinning one flux at the north pole and moving another away increases $\Delta M$ as the anyons separate, shown by the examples in Eq.(\ref{eq_l_middle}). If the two fluxes are added at the north and south poles, the electron occupation remains symmetric as in Eq.(\ref{eq_l_NS}), resulting in $L_z=0$ and $\Delta M_{max} = L_{max}$
\begin{align}
    \label{eq_l_NN}   & \textsubring{0}\textsubring{0}100100...1001\\
    \label{eq_l_middle}  & \textsubring{0}10\textsubring{0}0100100...1001 \ \ \textsubring{0}10010\textsubring{0}01...001 \ \ ... \\
    \label{eq_l_NS}    & \textsubring{0}100100...1001\textsubring{0}
\end{align}

\begin{figure}
\begin{center}
\includegraphics[width=\linewidth]{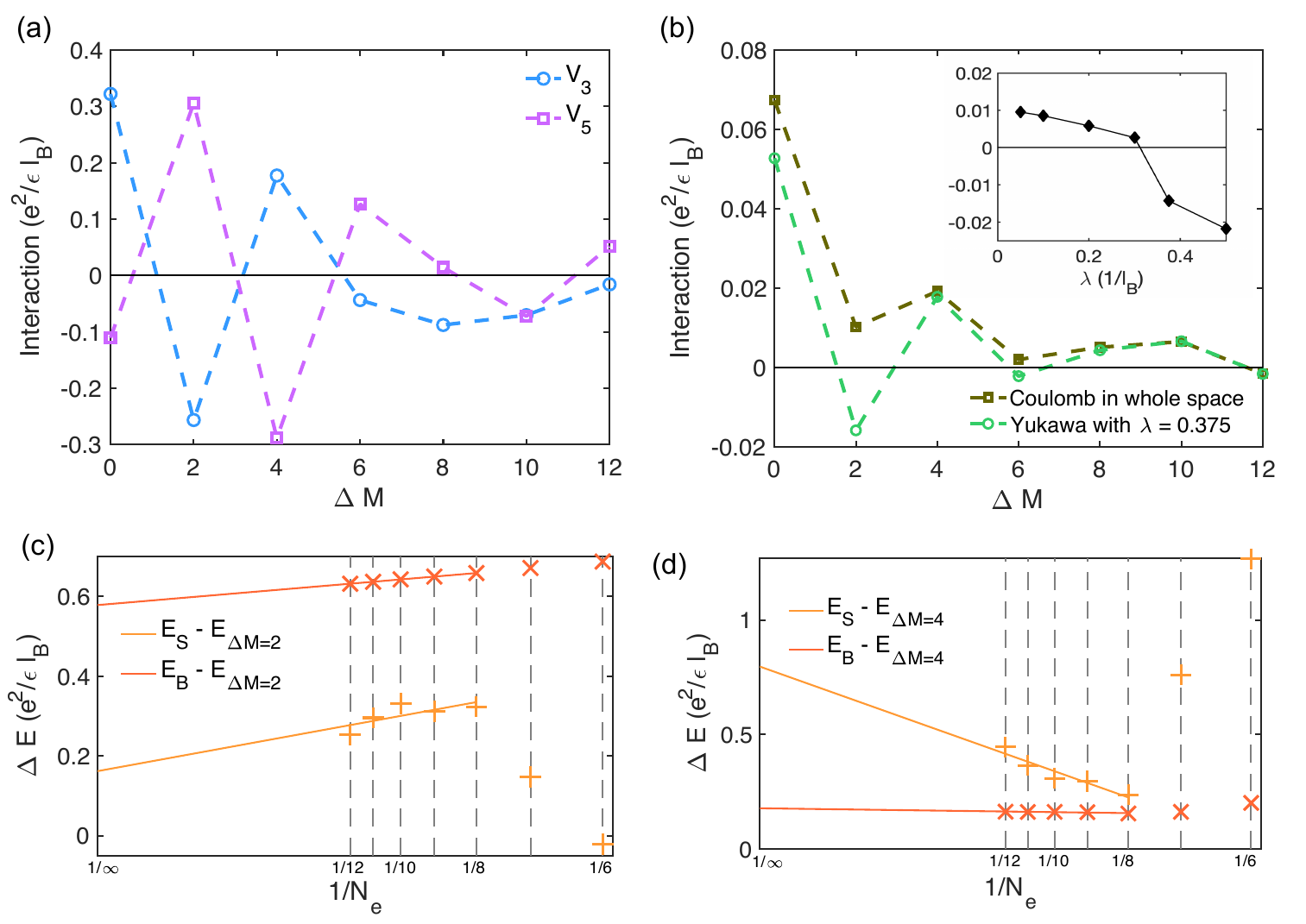}
\caption{(a) Interaction between two anyons with model electron-electron interaction $\hat{V}_3$ (blue circles) and $\hat{V}_5$ (purple squares)  in $\mathcal{H_L}$; (b) Effective interaction between two anyons induced by bare electron-electron interactions: Coulomb (olive squares) and Yukawa with $\lambda = 0.375/l_B$ (green circles), inset: energy differences between the bunching state $E_{\Delta M =2}$ and the most separated state $E_S$ ($\Delta M =12$ for the largest system size $N_e =12$ we computed) against $\lambda$. (c, d) Finite-size scaling of the energy differences between the most separated state $E_S$ (orange crosses), or the bound state $E_B$ when $\Delta M =0$ (yellow pluses), and the bunching state $E_{\Delta M =2}$ with $\hat{V}_3$ (c) and $\hat{V}_5$ (d).}
\label{fig_l_2qh}
\end{center}
\end{figure}

\subsection{Dynamics of Two Laughlin 1/3 Anyons}
We first study the interaction between two Laughlin 1/3 anyons under several leading two-body PPs using the model Hamiltonian 
\begin{equation}
    \hat{H}_{\text{int}} = t \hat{V}_1 + \hat{V}_{2k+1}, \ k = 1, 2, 3, ...
    \label{eq_model_H}
\end{equation}
with $t \gg 1$, projecting the studied Hilbert space to $\mathcal{H_L}$ (null space of $\hat{V}_1$) where the elementary particles are the Laughlin $1/3$ anyons, undressed from neutral excitations outside of $\mathcal{H_L}$. In our calculations, $t=100$ is typically set, which is sufficient to numerically enforce the projection, shown by a finite energy gap between the quasihole states and all the other states above them in the spectrum. In fact, we find that a value of $t$ just one order of magnitude larger than the coefficient of the additional e-e interaction term already ensures an effective separation between the quasihole sector and the higher-energy continuum.

Without $\hat{V}_{2k+1}$, these anyons are massless and non-interacting as an ideal gas in $\mathcal{H_L}$. After adding a relatively small $\hat{V}_{2k+1}$, anyons will gain mass and start to interact. Figure \ref{fig_l_2qh} (a) illustrates the effective interaction between two Laughlin $1/3$ anyons as a function of $\Delta M$ under the influence of the model electron-electron interactions $\hat{V}_3$ and $\hat{V}_5$. The interaction energy is defined as the total energy subtracted by the ground-state energy and the creation energies of all the quasiholes.
\begin{equation}
    E_{\text{int}}=E_{\text{total}}-E_\text{g} - N_{qh}E_{\text{qh}}
    \label{eq_effectiveINT}
\end{equation}
$E_g$ stands for the ground state energy and $E_{qh}$ is the self-energy of one quasihole (energy cost to create one quasihole). To compute $E_{qh}$, we introduce one additional flux into the system while keeping the number of electrons fixed, and subtract the resulting ground-state energy from $E_g$. This defines the creation energy of a single, well-isolated quasihole, independent of any angular momentum, spatial configuration, or presence of other quasiholes.

The interacting energies between two anyons are negative at a relatively small separation under both $\hat{V}_3$ and $\hat{V}_5$ electron-electron interactions in $\mathcal{H_L}$, shown in Figure \ref{fig_l_2qh}. Consequently, the two anyons will combine to form molecule-like structures with $\Delta M =2$ or $4$ being the lowest energy states, respectively. Finite-size scalings in (c, d) demonstrate that the key features, such as the formation and stability of the bound anyon cluster state, are robust and converge well when increasing system size. While a detailed understanding of the effective interaction's origin remains to be established, our studies and several indirect arguments in the literature support the relevance to the non-trivial statistics and real-space density oscillations of the anyons \cite{laughlin1988statistics, guangyue2024oscillation, Trung2023spin-statistics}. A deeper investigation into the precise roles of these features is an open question for future work.

\begin{figure*}[ht]
    \centering
\includegraphics[width=1\textwidth]{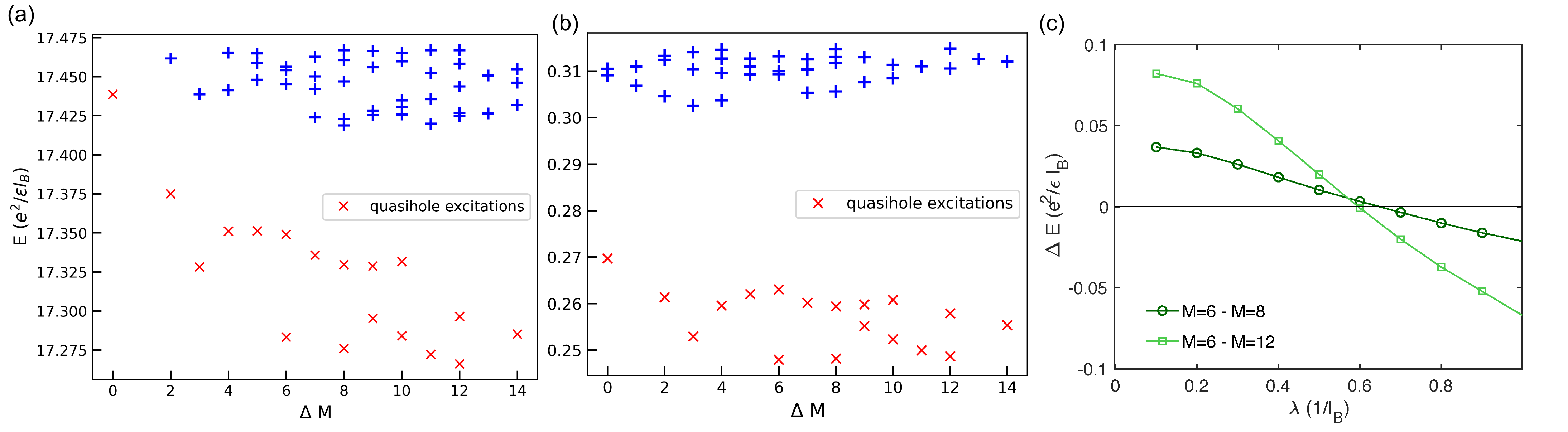}
    \caption{Energy spectrum of three-quasihole Laughlin 1/3 state with electron-electron (a) Coulomb interaction (b) Yukawa interaction when $\lambda=1/l_B$, both in the LLL; (c) Energy differences between the $\Delta M=6$ state and its competing ones-$\Delta M=8$ (olive circles) and $\Delta M=12$ states (green squares)-plotted as a function of $\lambda$, both exhibiting monotonic behaviours. Red crosses indicate quasihole states, blue pluses denote other excitations outside the QH manifold. The system size is $N_e =10$.}
    \label{fig_l3qh_real}
\end{figure*}

Building on the study with model Hamiltonians, we further explore the impact of realistic e-e interactions on two-anyon clusters. Forming an anyon molecule requires a relatively large coefficient for $\hat{V}_1$ and the dominance of $\hat{V}_{2k+1}$ with small $k$ based on the results of model Hamiltonians. We hence examine the anyon dynamics under the short-range Yukawa interaction that is commonly used to incorporate the screening effect theoretically:
\begin{equation}
    V (\mathbf{r}) = \frac{e^{-\lambda |\mathbf{r}|}}{|\mathbf{r}|}, \
    V (\mathbf{q}) = \frac{1}{\sqrt{\lambda^2 + |\mathbf{q}|^2}}
    \label{eq_yukawa}
\end{equation}
where $\lambda$ is the inverse of coherent length in units of $1/l_B$. Two Laughlin $1/3$ anyons tend to separate with Coulomb interaction, shown in Figure \ref{fig_l_2qh} (b). As $\lambda$ increases, the lowest energy state of the two-anyon cluster undergoes a monotonic transition from a deconfined state to a bunching state with $\Delta M = 2$, such as the green dashed line manifested where the effective anyon interaction is induced by $\lambda=0.375$ Yukawa interaction. The inset of Figure \ref{fig_l_2qh} (b) shows the monotonic trend of this transition, where the energy differences between the bunching $\Delta M = 2$ state and the separated state decrease gradually as $\lambda$ increases, changing from positive to negative. Since screening uniformly reduces the overall energy scale, we normalize the pseudopotential decomposition by fixing the coefficient of $\hat{V}_3$ to unity, and assign the remaining pseudopotential coefficients according to their original relative ratios. This ensures that all quantities such as energy gaps and total energies remain comparable across different values of $\lambda$ (and Coulomb).

The two-anyon “molecule” introduced by the short-range electronic interactions provides a plausible explanation for the unexpected results in quantum point contact experiments, where the effective charge carrier shows a temperature dependence, transitioning from $2e/3$ at low temperatures ($\sim10mK$) to $e/3$ at higher temperatures ($\sim100mK$) \cite{efcharge-PRL2009shot, efcharge-PRB2010, efcharge-PRL2011,efcharge-NP2017}. In shot noise experiments, Coulomb screening will arise from the 2D dielectric environment, nearby metallic gates, and quantum point contact potentials, which will distort electric fields, suppress long-range interactions, and enhance short-range effects \cite{exp_shot_noise_2007, exp_shot_noise_ajit2013, exp_shot_noise_2020}. Based on the above discussion, at near absolute zero temperature with the effect of short-range interaction, the effective charge carriers in the quantum fluid are bunching “molecules” with $e^* = 2e/3$ formed by two Laughlin 1/3 anyons. Thermal fluctuations will gradually break apart these “molecules” as temperature rises, and the charge carrier will completely change back to the single Laughlin 1/3 anyons with $e^* = e/3$ in the end. A quantitative comparison between the binding energy of the anyon “molecule” under Yukawa interaction with $\lambda = 0.375 / l_B$ ($10^{-24} \sim 10^{-25} \text{J}$) and the thermal energy at $T \sim 10-100 \text{mK}$ confirms that their scales are comparable, supporting this scenario.

\subsection{Interaction Within Three Laughlin 1/3 Anyon Cluster}
To prepare for our later discussions about bulk tunneling experiments—which is essential for understanding many-anyon FQH fluids in realistic scenarios—we investigate the interactions within a cluster of three Laughlin 1/3 anyons. This configuration naturally arises when one electron is removed from the Laughlin 1/3 state. Although the three-anyon case is more complex than the two-anyon scenario, it follows a similar trend. The total relative angular momentum reflects the extent of the anyon cluster. The ground states of both $\hat{V}_3$ and $\hat{V}_5$ in $\mathcal{H_L}$ correspond to a three-anyon cluster with total relative angular momentum $\Delta M = 6$. The interacting energy as a function of the total relative angular momentum $\Delta M$ within this anyon cluster is provided in the Supplementary Materials, where finite-size scaling of the interaction energy gaps between $\Delta M = 6$ and its competing states in both cases confirms the stability of $\Delta M = 6$ state in the thermodynamic limit. 

We studied the three-anyon spectrum for both Coulomb and Yukawa potentials with varying screening length $\lambda$. In all panels of Figure \ref{fig_l3qh_real}, red crosses mark states in the quasihole CHS, identified via a large-t projection using the model Hamiltonian $\hat{H} = t \hat{V}_1 + \hat{V}_{\text{int}}$, where there exists a finite energy gap between the quasihole manifold and the other higher-energy states. They are tracked as $t\rightarrow0$ toward the physical interaction, during which the structure of the QH manifold -- particularly its state counting -- remains unchanged because of its topological nature \cite{HuiLi2008entanglement}. Blue plus symbols denote all the other states outside the quasihole manifold, which are not the subject of this paper and won't be further identified. 

With LLL Coulomb interaction, the three-anyon cluster's lowest energy state is $\Delta M =12$, indicating a loosely bound structure. However, the configuration transitions to a more compact cluster with $\Delta M = 6$ as $\lambda$ in the Yukawa potential increases, corresponding to shorter-range electron-electron interactions. This transition is shown by depicting the energy difference between $\Delta M = 6$ ($E_{\Delta M = 6}$) and its competing states $\Delta M = 8$ or $\Delta M = 12$ as a function of $\lambda$ in Figure \ref{fig_l3qh_real} (c). $\Delta M = 6$ gradually becomes the lowest energy state as $\lambda$ increases. This aligns with our earlier finding in two-anyon cases that sufficiently short-range interactions dominated by $\hat V_1$ can replicate the effects of the model Hamiltonian Eq.(\ref{eq_model_H}) on the effective anyon interaction. It can also be seen from the whole spectrum of the system that the quasihole excitations (red crosses) gradually separate from other states (blue pluses) when increasing the $\lambda$, since during this process, $\hat{V}_1$ gradually becomes more dominant. We estimate the size of the anyon cluster using the first and second moments of their density distribution \cite{Papic2014size}. The estimated spatial radius of the lowest-energy anyon cluster is $40\sim 60 \text{nm}$. Density for the system under Coulomb e-e interaction indicates long-range correlations and oscillatory structures, while for Yukawa interaction, it has a prominent central peak and mild, smooth oscillations extending toward the edges, reflecting strong short-range binding ~\cite{SM}.

\section{Interactions within a cluster of Moore-Read Anyons}
The Moore-Read (MR) FQH fluid has attracted much greater attention because of its non-Abelian nature and potential applications in quantum computations. Unlike the case of the Laughlin state, two quasiholes, each with a $e/4$ charge, will be created by one single flux insertion in the MR ground state because of the flux fractionalization in non-Abelian phases. When fused, the two-quasihole excitation with $e/2$ charge collectively carries topological quantum numbers “$1$” or “$\psi$”, exhibiting a statistical phase $0$ or $\pi$ respectively \cite{S.Simon2009MR}. The topological numbers of these two kinds of excitations come from the fact that they follow the Ising model in conformal field theory (CFT) with energetically degenerate fusion channels “$1$” and “$\psi$”, corresponding to the U(1) boson and Majorana fermion chiral edge modes \cite{MooreRead1991, nayak1996statistics, nayak2008NA, moller2011oddeven}. For the same reason, a single $e/4$ quasihole excitation corresponds to the index “$\sigma$” in the Ising CFT model. In a general MR phase, the “$1$” and “$\psi$” types of quasihole excitations coexist, which is the reason for non-Abelian statistics. The degeneracy of a MR phase containing $N_{qh}$ quasiholes is $2^{N_{qh}/2-1}$. 

\subsection{Dynamics of Two MR Anyons}
The root configuration for the MR ground state is $110011...0011$, satisfying the $(4,2)$ admissible rule and the condition $N_o = 2 N_e -2$. We begin with the case of two quasiholes (one extra flux quanta) where the topological number of the system can only be either 1 or $\psi$, depending on whether the number of electrons ($N_e$) in the finite system is even or odd. These two types of two-quasihole excitations are illustrated by the following root configurations:
\begin{align}
&\doublesubring{\blankchar}01100110011...110011 &\text{$1$-type, even $N_e$} 
    \label{eq_root_even}\\    
   &\doublesubring{1}00110011...110011 &\text{ $\psi$-type, odd $N_e$}
    \label{eq_root_odd}
\end{align}
Each empty circle represents one quasihole with $e/4$ charge. The parity of the number of electrons in between the pair of $e/4$ MR anyons is the same as the parity of the whole system's $N_e$.

To compute the self-energy of a single $e/4$ quasihole in the MR phase, we fix the two quasiholes at the north and south poles of the sphere by Jack polynomials as $1010...10101$ and extract the energy from the ground-state energy $E_g$ with the same electron number. In the large-system limit, where their mutual interaction vanishes, the residual energy should approach twice the quasihole self-energy. More detailed explanations and discussion are provided in the Supplementary Material.
 
\begin{figure}
    \centering
\includegraphics[width=0.5\textwidth]{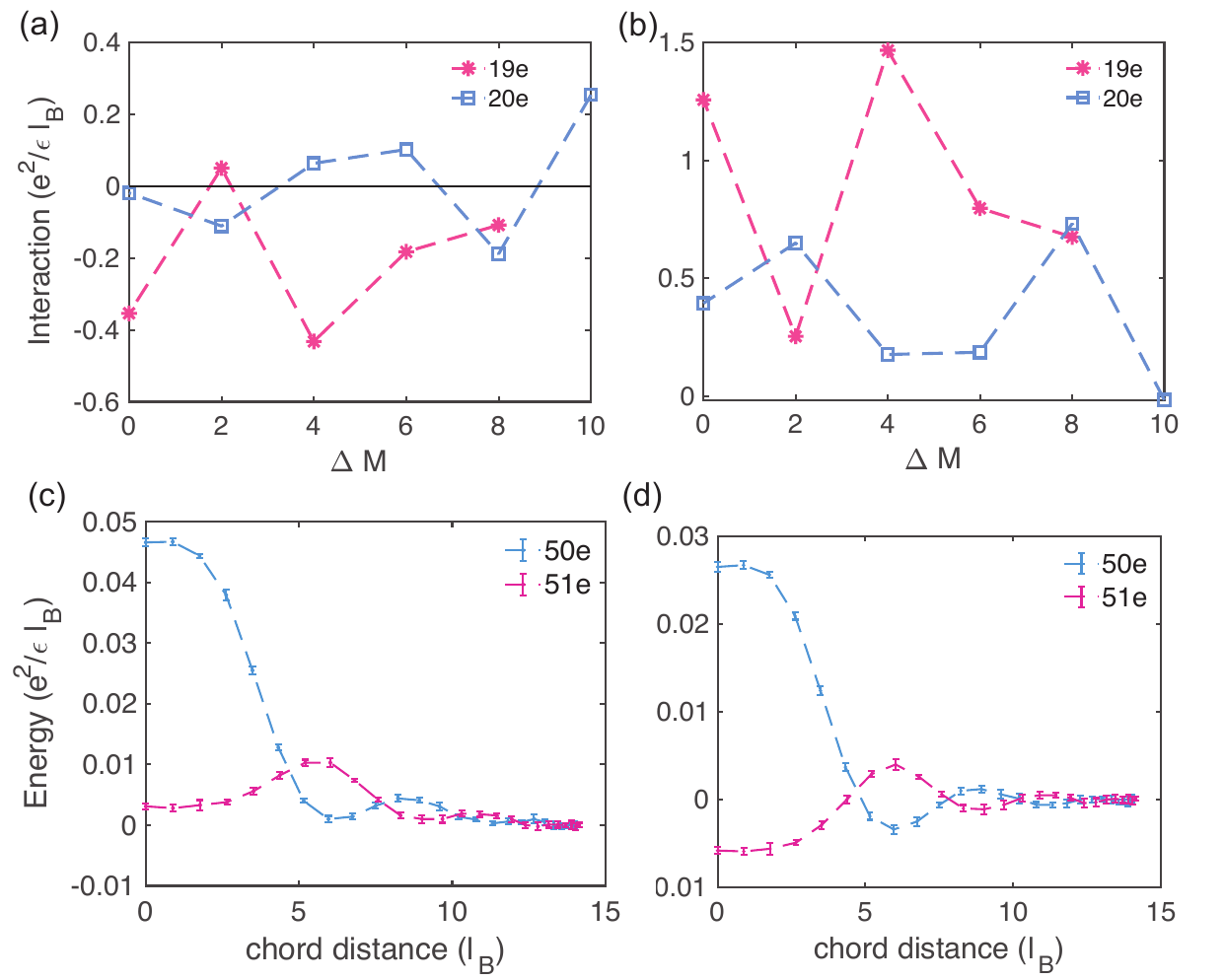}
    \caption{Effective interaction of two MR quasiholes with different electron-electron interactions. (a) $\hat{V}_1$ and (b) $\hat{V}_3$ within $N_e = 19$ (odd, pink stars) and $N_e = 20$ (even, blue squares) systems, calculated from ED; (c) Coulomb interaction and (d) Yukawa interaction when $\lambda = 0.5 /l_B$ in systems of $N_e = 51$ (odd, pink) and $N_e = 50$ (even, blue), computed by Monte Carlo with error bars shown.}
    \label{fig_mr_2qh}
\end{figure}

First, the effective interaction between two MR anyons under the model e-e interaction is studied. Haldane pseudopotentials $\hat{V}_1$ and $\hat{V}_3$ are diagonalized within the null space of the MR model Hamiltonian, defined by a special three-body interaction\cite{greiter1992_3-3bdy, simon2007projection} and obtained by a similar strategy as Laughlin's case Eq.(\ref{eq_model_H}) numerically, with the presence of two quasiholes. The effective anyon interactions defined by Eq.(\ref{eq_effectiveINT}) with $N_{qh}=2$ are shown in Figure \ref{fig_mr_2qh} (a) and (b), plotted against their relative angular momentum $\Delta M$. It depends on the parity of the system, as a manifestation of the odd-even effect in the MR phase \cite{S.Simon2009MR}. Since the resulting dynamics have distinct behaviors under different PPs by which the e-e interaction can always be decomposed, it suggests that tailored electron-electron interactions could exploit this distinction to manipulate the two types of MR anyons. For instance, the e-e interaction dominated by $\hat{V}_1$ induces the effective interactions for $\psi$-type anyons to bind together while favoring a comparatively large structure of $1$-types. Conversely, to introduce an effectively repulsive interaction between the $\psi$-type anyons for non-abelian braiding (therefore the possibility of topological quantum computation), e-e interaction dominated by $\hat{V}_3$ is preferred.

\begin{figure*}
    \centering
\includegraphics[width=1\textwidth]{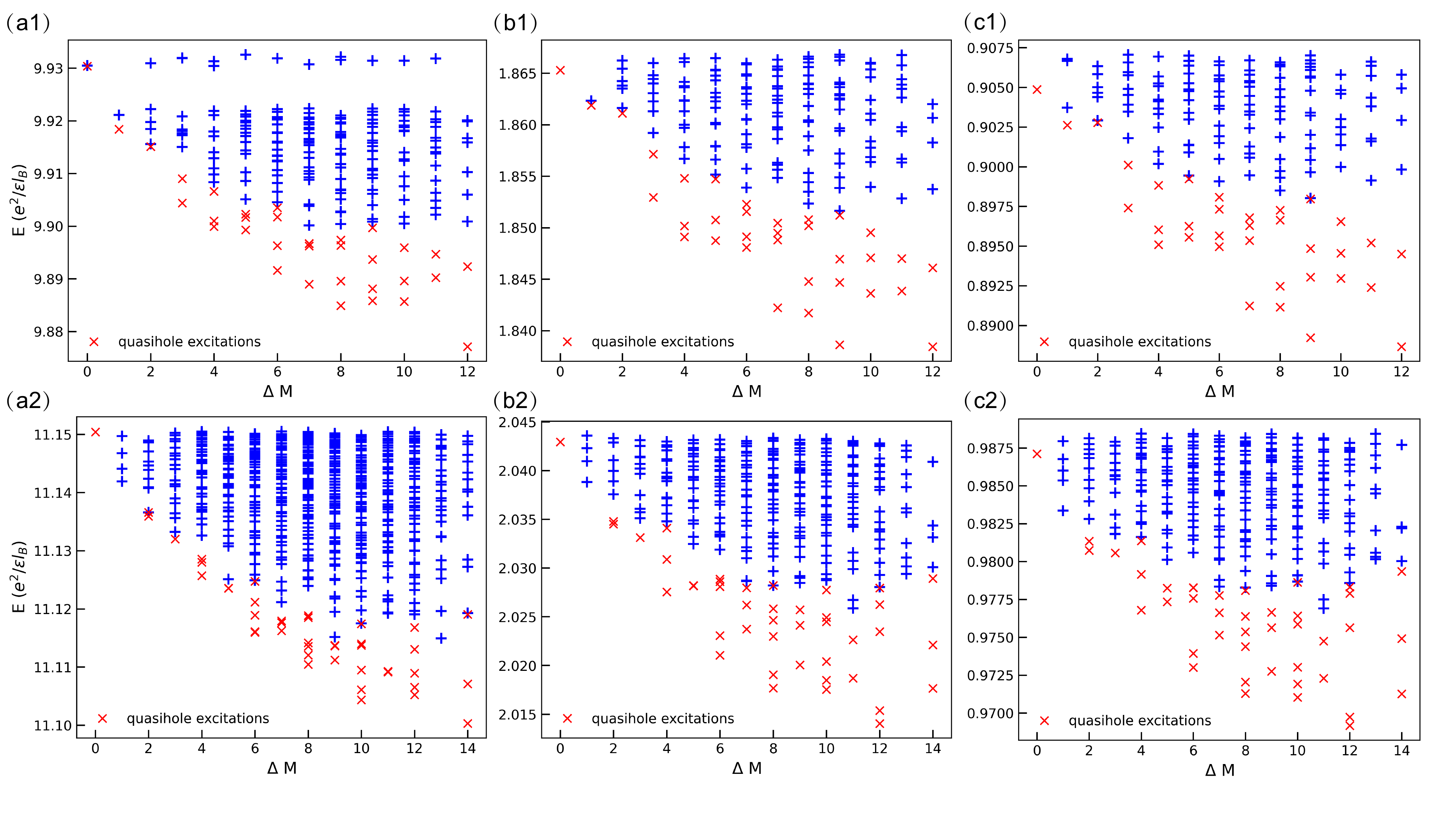}
    \caption{Spectrum of MR with four quasiholes under various realistic interactions, with odd (here $N_e=13$, first row) and even (here $N_e=14$, second row) numbers of electrons. (a) Coulomb interaction; (b) Yukawa with $\lambda=0.5$; (c) Yukawa with $\lambda = 1$.}
    \label{fig_mr4qh_spectrum}
\end{figure*}

Effective interactions between two MR anyons under realistic electron-electron interactions are more challenging to study due to the strong finite-size effect, which requires larger system sizes that are difficult to achieve by exact diagonalization. To overcome this limitation, we employ the Monte Carlo simulation to explore larger systems. The energies when two localized quasiholes are gradually separated in systems with $N_e=50$ (even) and $N_e=51$ (odd) are computed using the Pfaffian wavefunction by fixing one anyon at the north pole and moving the other apart. The final average stable value, corresponding to the regime where the anyons are sufficiently separated and no longer interact, is taken as the background energy and subtracted from all data points. 

The resultant effective interactions between two localized $e/4$ MR quasiholes are plotted as a function of their chord distance. The odd-even effect remains prominent, with nearly opposite dynamical characteristics observed between systems with the two different parities. Under Coulomb interaction shown in Figure \ref{fig_mr_2qh} (c), the effective interactions are repulsive at all distances for systems with $N_e$ of both parities. With shorter-range Yukawa interactions example $\lambda = 0.5/l_B$ in Figure \ref{fig_mr_2qh} (d)-the effective interaction for $\psi$-types is attractive at small separations. But it remains repulsive for $1$-types with the same distances, then develops to be attractive at a comparatively large separation. This behaviour indicates that short-range interactions energetically favor tightly bound pairs of $e/4$ MR anyons in odd $N_e$ systems, while promoting spatially extended, loose configurations of them in even $N_e$ systems. It means that the $\psi$ fusion channel is energetically more favorable under the short-range interaction than the $1$ channel when considering the fusion of two “$\sigma$” anyons with $e/4$ charges. These results further support the feasibility of engineering bare electron-electron interactions to selectively separate and control the two types of fusion channels of the $e/4$ MR anyon and the consequent $e/2$ excitations.

\subsection{Interaction Within Four MR Anyon Clusters}
We now look at a cluster of four MR quasiholes—each carrying a $e/4$ charge—which will be useful in Sec .~\ref{tunneling}. A four-MR-anyon cluster emerges upon the removal of a single electron from the ground state. The resulting system exhibits a degeneracy of $2^{4/2}-1=2$, where $1$- and $\psi$-types could coexist, adding significant complexity to the analysis. Specifically, the two pairs of quasiholes can only be two 1-types or two $\psi$-types in even $N_e$ systems, whereas the configuration necessarily consists of one 1-type and one $\psi$-type in odd $N_e$ systems \cite{trung2025longrange}. The energy spectrum of this system under model electron-electron interactions $\hat{V}_1$, $\hat{V}_3$, and $\hat{V}_5$ in the MR null space (in supplementary \cite{SM}) reveals two key features: a persistent odd-even effect in systems with four MR quasiholes and a preference of loose anyon configurations characterized by large relative angular momentum $\Delta M$ within the four-anyon cluster. 

To extend our investigation to more realistic scenarios, given that MR states are proposed to be stabilized in the 1LL, we studied the e-e Coulomb and Yukawa potentials in the 1LL shown in Figure \ref{fig_mr4qh_spectrum}. In contrast to Laughlin's case, the ground states of MR systems with four anyons consistently exhibit a loose structure and remain largely insensitive to variations in electron-electron interactions. For the system with $N_e =13$, the lowest energy state is the one with relative angular momentum $\Delta M =12$, the maximum allowed $\Delta M$ value in this finite system. In the $N_e =14$ case, the four-anyon cluster attains the largest possible value of $\Delta M =14$ under the 1LL electron-electron Coulomb interaction, and changes to $\Delta M =12$ for both Yukawa interactions we computed with $\lambda = 0.5$ and $1$. The results are not conclusive due to the limited system size we can reach.  Another observation is that the quasihole excitation spectrum maintains a comparatively robust, model-independent structure, unchanged not only across the parameter regimes studied among the realistic interactions, but also across the realistic interaction and the model Hamiltonian inside the MR CHS, whose spectrum can be further found in Supplementary Materials. Using the first and second moments of the density distribution, the estimated real-space radius of the four-quasihole cluster is $100\sim 140 \text{nm}$ for the lowest energy state as $\Delta M = 12$ in $N_e = 15$ (odd) system. And for the $\Delta M = 14$ anyon cluster in $N_e = 16$ (even) system, it is $75\sim 110 \text{nm}$. The spatial extent of the anyon clusters with the lowest energy in the MR phase exhibits a difference between odd and even electron systems, suggesting a parity effect in the cluster configuration in the finite
systems. And the predicted spatial extents are much larger compared to Laughlin’s cases ~\cite{SM}.

\section{Bulk tunneling experiments}\label{tunneling}
The effective interactions within anyon clusters have direct experimental relevance in recently developed STM measurements, whose detailed setups and implementations, both theory and experiment, have been discussed in several recent works \cite{exp_STM_Papic2018, exp_STM_highReso, exp_STM_coissard2022imaging}. In such experiments, single-electron tunneling occurs when the electronic potential $V_t$ applied to the tip satisfies 
\begin{equation}
    E =eV_t \geq E_0 + E_z + E_{xy}
\end{equation}
where $E_0$ is the work function of the 2DEG, $E_z$ accounts for the finite-thickness effect \cite{DasSarma2008FiniteWidth}, and $E_{xy}$ represents the energy scale of the strongly correlated 2D system studied here. After the bulk tunneling, a few-anyon cluster with total positive charge of $+e$ will be left behind. The number of anyons in this cluster is determined by the specific FQH phase (e.g. $3$ for Laughlin 1/3 and $4$ for MR). Due to the non-trivial inter-anyonic interactions, as we have shown in the previous sections, the anyon cluster could be in different forms of “bound states" depending on the effective anyon interaction, the size of the STM tip, and the bias voltage applied to the tip. In particular, the effective interactions and the consequent structures of the few-anyon clusters influence the experimental outcomes, especially for the low-lying parts, which are mainly contributed by the quasihole states. It makes such measurements potentially very useful for understanding the dynamics of anyon clusters.

\begin{figure}
    \centering
    \includegraphics[width=1\linewidth]{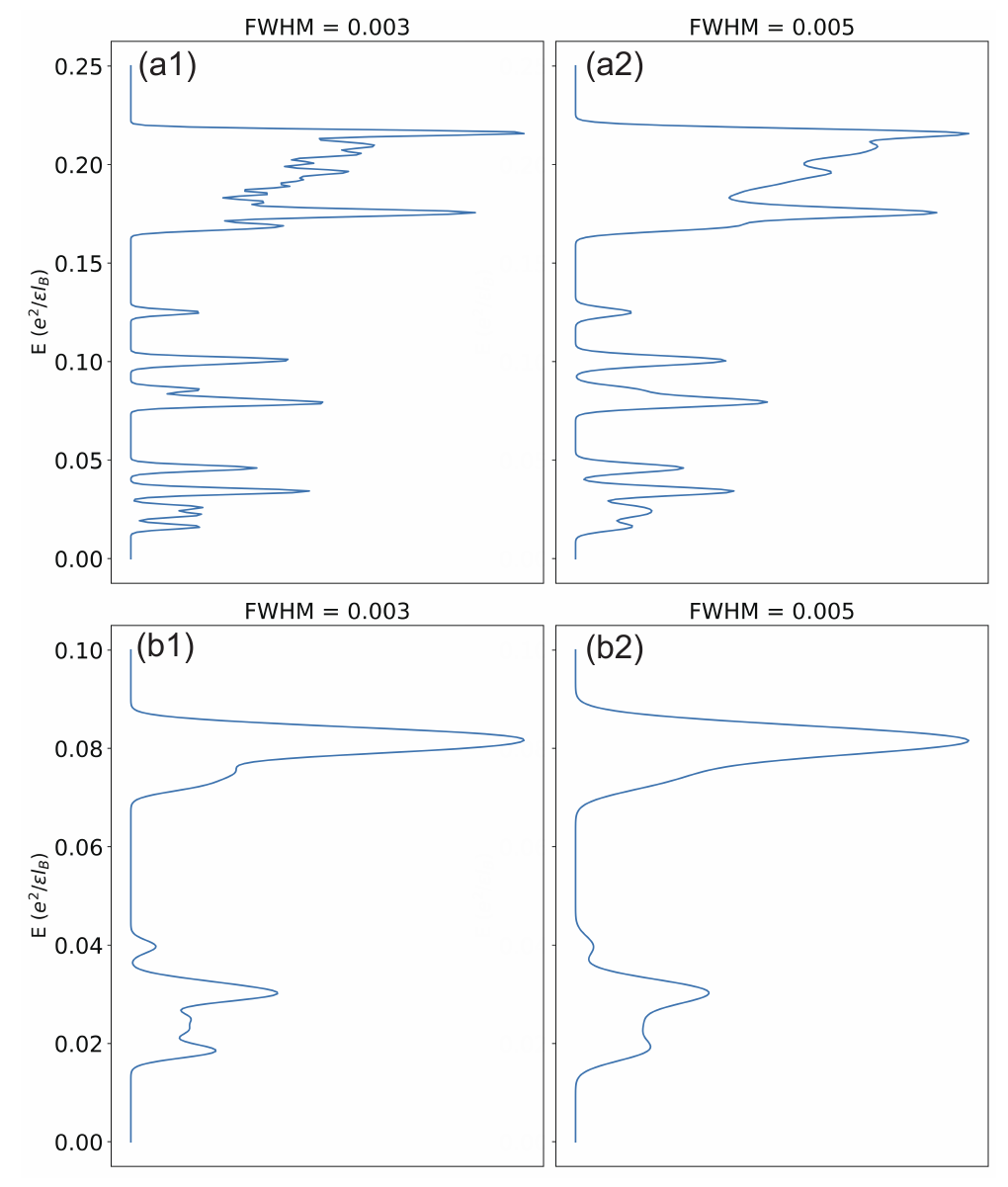}
    \caption{Possible low-energy LDOS for Laughlin 1/3 state with $N_e = 10$ at FWHM = $0.003\text{(first panel)}/0.005\text{(second panel)}\ e^2/\epsilon l_B$. The entire background energies are subtracted in the subfigures since only the energy differences matter here. (a) Coulomb interaction, (b) Yukawa interaction when $\lambda = 1$, both in the LLL.}
    \label{fig_stm_laughlin}
\end{figure}

The local density of states (LDOS) is a key observable in STM measurements. High-resolution tunneling can reveal a clear gap of approximately $ 0.1 meV$, which corresponds to $\Delta E = 0.002 \sim 0.006\ e^2/\epsilon l_B$ whose exact value depends on the specific dielectric constant $\epsilon$ and the magnetic field $B$ under detection. According to this range, we simulated the possible LDOS for the 3-anyon Laughlin 1/3 states and 4-anyon MR states, focusing on the low-lying parts of the states mainly from the quasihole manifold. It is found that the variations in effective anyon interactions—induced by different bare e-e interactions—lead to distinct LDOS patterns, including differences in the numbers and positions of the peaks. 

In this work, we follow the standard STM approximation where the tunneling amplitude is assumed to be constant and featureless, so the measured tunneling conductance is proportional to the local density of states (LDOS) of the sample. This approach omits the energy-dependent tunneling matrix element computed in more microscopic treatments such as Bardeen’s formalism, which derives from Fermi’s golden rule\cite{chen2021introduction}. While the detailed tunneling amplitudes may vary with the STM tip geometry, material-specific parameters, or symmetry constraints, our analysis aims to capture the qualitative LDOS features arising from the internal structure of few-anyon bound states. A more refined modeling of the tunneling matrix elements will be pursued in future work.

Our analysis shows that the low-lying tunneling spectrum is rich, affected by the effective anyon interaction and the internal structure of the anyon cluster, both in the Laughlin and Moore-Read phases. We only present a preliminary study here, assuming the bias voltage of the STM tip does not alter the anyon clusters' densities. This may not be the case given that the one-body potential for an electron to tunnel highly depends on the geometry of the tip, the bias voltage, as well as the distance between the tip and the Hall manifold \cite{trung2025longrange}. A more detailed study with explicit experimental parameters will be presented in the future.

\subsection{LDOS Measurement for Laughlin 1/3 Phase}

The low-energy spectrum for a system containing three Laughlin 1/3 anyons with realistic electron-electron interactions has already been shown in Figure \ref{fig_l3qh_real} (a), where the most tightly bound anyon cluster state ($\Delta M = 0$ state) lies at significantly higher energy than the other low-lying excitations for relatively long-range e-e interactions such as LLL Coulomb. As the screening parameter $\lambda$ increases, all quasihole excitations shift downward and gradually separate from the high-energy continuum. The $\Delta M = 0$ state eventually becomes distinguishable from the gapped excitations but remains the highest in energy among all quasihole states. The simulated LDOS for $FWHM=0.003$ and $0.005$ under e-e Coulomb interaction and $\lambda=0.5$ Yukawa interaction, both in LLL, are shown in Figure \ref{fig_stm_laughlin}. The LDOS results for different interactions are clearly distinguishable as in Figure \ref{fig_stm_laughlin}. This indicates that LDOS can serve as a fingerprint for characterising the effective interactions among anyons.

\begin{figure*}
    \centering
    \includegraphics[width=1\linewidth]{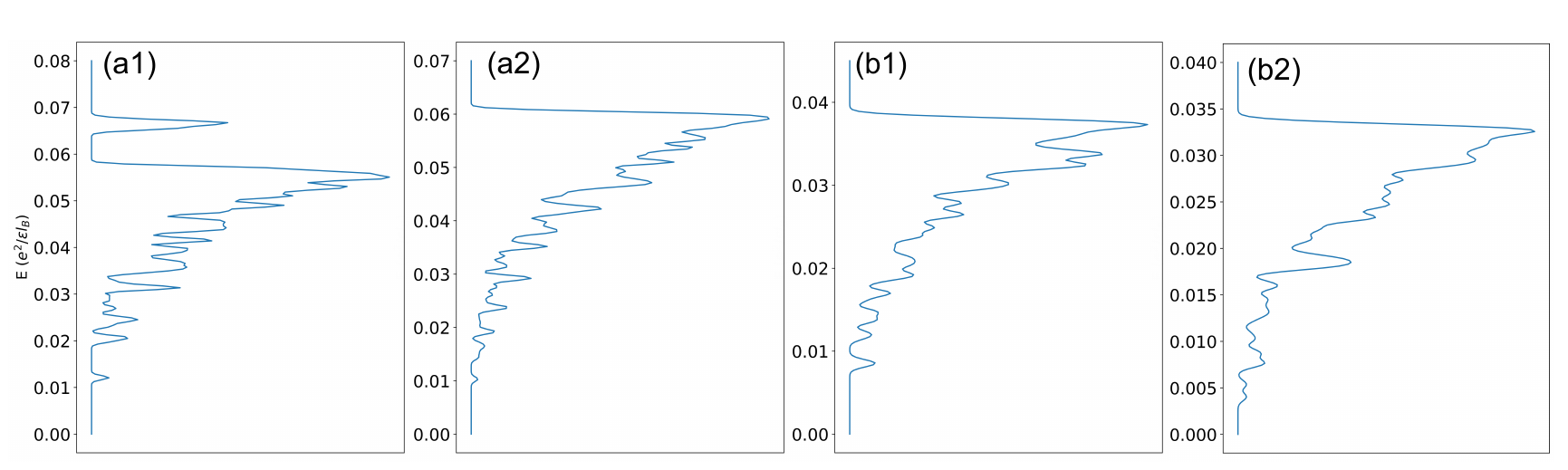}
    \caption{LDOS of the MR system containing 4 anyons at FWHM = $0.001\ e^2/\epsilon l_B$ for: 1LL Coulomb interaction with (a1) $N_e = 13$ and (a2) $N_e = 14$; 1LL Yukawa interaction when $\lambda = 0.5$ with (b1) $N_e = 13$ and (b2) $N_e = 14$.}
    \label{fig_stm_mr}
\end{figure*}

\subsection{LDOS Measurement for Non-Abelian Moore-Read Phase}

In the MR phases, the $\Delta M = 0$ state is even more deeply buried in the continuum of gapped excitations, regardless of the e-e interaction being long-range 1LL Coulomb or comparatively short-range 1LL Yukawa potentials shown in Figure \ref{fig_mr4qh_spectrum}. Across all simulated LDOS results for both odd and even $N_e$ systems under various e-e interactions, the most experimentally detectable lowest peaks under the previous assumption correspond to states having very large relative angular momentum within the MR anyon cluster. However, depending on the strength of the bias voltage and the details of the STM tip, it does not necessarily correspond to the lowest energy state. 

Notably, even for MR phases where the energy spectra appear similar under various realistic electron-electron interactions, as we mentioned in the last section, the LDOS remains sensitive to these variations, manifested from the strengths and locations of the density peaks in Figure \ref{fig_stm_mr}. This underscores STM's potential as a tool for detecting and verifying effective anyon interactions in both Abelian and non-Abelian FQH systems. However, compared to the Laughlin state—where peaks are well separated even at lower resolutions—MR systems require measurements of higher resolution (for example, FWHM = $0.001\ e^2/\epsilon l_B$ in Figure \ref{fig_stm_mr}). This is understandable since removing one electron creates three Laughlin quasiholes, but four MR quasiholes, and the latter leads to many more configurations with similar energies.

We emphasize that while detailed peak positions and splittings in the LDOS depend on the microscopic form of the interaction, certain qualitative features are universal. For example, in the Laughlin case, the emergence of two dominant peak clusters from the 3-quasihole spectrum is present across different interaction models and FWHM values, where the lower cluster contributes to the quasihole excitations. In the Moore–Read case, the LDOS shows a clear distinction between even and odd electron number sectors, a direct consequence of the underlying non-Abelian fusion rules. These features are expected to survive under reasonable experimental conditions and provide a possible path toward distinguishing topological signatures from non-universal effects in STM experiments.
 



\section{Conclusion}
In this paper, we investigated the effective interactions of anyons in Laughlin and Moore-Read FQH phases derived from diverse electron-electron interactions, both model and realistic ones. The effective interaction within anyons is an interpolation between fermions' repulsive and bosons' attractive force, exhibiting various and unusual behaviors depending on the detailed forms of the electron-electron interaction in the 2DEG. The recently developed STM experiments in FQH systems could be a new platform to explore these properties of anyons. As we show, many bound anyon states with different internal structures will be involved in the low-lying energy parts of the tunneling spectrum. We point out that effective anyon interactions are important factors in understanding this part of the results in STM tunneling experiments. Since anyons can be considered the elementary particles for the corresponding FQH phases and are potentially useful for the storage and processing of quantum information, the interplay between their effective interactions and the correlated topological phases should be considered as a significant feature when investigating their properties both theoretically and for future applications.

\paragraph*{Note.}
While this manuscript was under review, a related preprint appeared ~\cite{gattu2025molecular} on the formation of anyon clusters (anyon “molecule”), readers may find it complementary.

\begin{acknowledgments}
Some of the numerics in this work are performed using the DiagHam package for which we would like to thank all the authors. This work is supported by the National Research Foundation, Singapore, under the NRF Fellowship Award (NRF-NRFF12-2020-005), Singapore Ministry of Education (MOE) Academic Research Fund Tier 3 Grant (No. MOE-MOET32023-0003) “Quantum Geometric Advantage”, and Singapore Ministry of Education (MOE) Academic Research Fund Tier 2 Grant (No. MOE-T2EP50124-0017).
\end{acknowledgments}

\nocite{MilovanovicRead1996,ReadGreen2000,RezayiHaldane2000, gattu2025molecular}

\bibliography{apssamp}

\clearpage
\onecolumngrid   

\begin{center}
  \textbf{\large Supplemental Material for\\[0.3em]
  “Dynamics of Anyon Clusters in Fractional Quantum Hall Fluids”}
\end{center}
\vspace{1em}

\setcounter{equation}{0}
\setcounter{figure}{0}
\setcounter{table}{0}
\setcounter{section}{0}
\renewcommand{\theequation}{S\arabic{equation}}
\renewcommand{\thefigure}{S\arabic{figure}}
\renewcommand{\thetable}{S\arabic{table}}

\section{Haldane Pseudopotentials}
Generally speaking, the matrix element of a rotationally invariant interaction $\hat{V}$ can be expanded in the coupled basis as:
\begin{align}
    \langle{s_1,s_2}| \hat{V} |{s_3, s_4} \rangle
    &= \sum_{L, L^\prime=0}^{2l} \sum_{M=-L}^L \sum_{M^\prime=-L^\prime}^{L^\prime} \langle{l_1,s_1; l_2, s_2}| L,M\rangle V_L \langle{L^\prime, M^\prime}|
    l_3,s_3; l_4, s_4\rangle \\
    V_L &= \langle{L,M}| \hat{V} |{L^\prime, M^\prime}\rangle = \delta_{L, L^\prime} \delta_{M, M^\prime} \langle{L,M}| \hat{V} |{L^\prime, M^\prime}\rangle \notag \\
    \langle{s_1,s_2}| \hat{V} |{s_3, s_4} \rangle
   & = \sum_{L=0}^{2l} \sum_{M=-L}^L \langle{l_1,s_1; l_2, s_2}| L,M\rangle
   \langle{L,M}| \hat{V} |{L, M}\rangle
   \langle{L, M}| l_3,s_3; l_4, s_4\rangle \notag
\end{align}
$\langle{l_1,s_1; l_2, s_2}| L, M\rangle$ (as well as $\langle{L, M}| l_3,s_3; l_4, s_4\rangle$) is the Clebsch-Gordan coefficient on the sphere when making the basis transformation of a two-particle state between single-particle angular momentum total angular momentum representation. Rotational invariance implies $V_L$ is independent of $M$, and the selection rules are
\[
M=s_1+s_2=s_3+s_4,\qquad
|\,l_1-l_2\,|\le L\le l_1+l_2 
\] 

On the sphere, the (pair) total angular momentum $L$ in the same LL with shell angular momentum $l$ is related to the relative angular momentum $m$ by $m \equiv 2l - L$. For identical spin-polarized fermions, $m$ must be odd (even for bosons). It is convenient to parameterize $\hat V$ by the Haldane pseudopotentials
(the pair energies in fixed relative-angular-momentum sectors):
\begin{equation}
c^{(n)}_{m} \;\equiv\;
\big\langle L=2l-m,\,M \big| \hat V \big|L=2l-m,\,M \big\rangle
\end{equation}
where $n$ is the Landau-level index (we suppress $n$ below since the LL is fixed). $c^{(n)}_m$ is the coefficient of so-called $m_{th}$ Haldane pseudopotential at nLL \cite{haldane1983hierarchy}. The RHS does not depend on $M$ because of the rotational invariance. Then $\hat V$ admits the spectral decomposition
\begin{align}
    \hat{V} 
    = \sum_{m=0}^{2l} c^{(n)}_m \hat{P}_{12}(m) 
\end{align}
Here we define the projector onto the two-particle subspace with given
relative angular momentum $m$ (equivalently, total $L=2l-m$) as:
\begin{equation}
\hat P_{12}(m)
\;=\;
\sum_{M=-(2l-m)}^{2l-m}
\big|2l-m,\,M\big\rangle\big\langle 2l-m,\,M\big|
\end{equation}
For fermions (bosons), only odd (even) $m$ terms contribute. Larger $m$ corresponds to larger average particle separations, so $V_m$ encodes progressively longer-range components of the interaction. The $m_{th}$ Haldane pseudopotential represents the energy for a pair of electrons to stay in the state with a specified relative angular momentum m. 

A specific point on the sphere can be indicated by its radius part $\mathbf{r}=R\ \mathbf{e}_r$ and angular part $\mathbf{\Omega} = (\theta,\phi)$ in the spherical coordinate. $\mathbf{\Omega}$ is tangential and does not depend on the radial part. $\theta \in [0, \pi], \phi \in [0, 2\pi]$. Introduce the spinor coordinates as:
\begin{equation}
    u = \cos{\frac{\theta}{2}} \exp{(\frac{i \phi}{2})},\ \ v = \sin{\frac{\theta}{2}} \exp{(-\frac{i \phi}{2})}
\end{equation}
The two-particle wavefunction for a pair of electrons with total angular momentum $L$ (namely, relative angular momentum $m=2S-L$) is denoted by:
\begin{equation}
    \psi_{(\alpha,\beta)}^{l,L}[u,v] = (u_1v_2-u_2v_1)^{2l-L}\ \Pi_{i=1,2} (\bar{\alpha} u_i + \bar{\beta} v_i)^L
\end{equation}
Normally we choose $(\alpha,\beta)=(1,0)$ for simplicity:
\begin{equation}
    \langle \mathbf{\Omega} |L,M \rangle = \psi_{(1,0)}^{(l,L)}[u,v] = (u_1v_2-u_2v_1)^{2l-L}\ u_1^L u_2^L
\end{equation}

The explicit computation of pseudopotential decomposition has several ways and here we give an example in computing it using the above mentioned set-ups. Some other detailed calculation can be found in other references. The formulation is general and can be applied to any rotationally invariant two-body interaction in a Landau level.

The chord distance between two particles can be written as:
\begin{equation}
    |\mathbf{r}_1-\mathbf{r}_2| = 2R|u_1v_2-u_2v_1|
\end{equation}
Yukawa potential (set the radius $R=1$ for simplicity since it is just a constant):
\begin{equation}
    V_{Y} = \frac{e^{-\lambda |\mathbf{r}_1-\mathbf{r}_2|}}{|\mathbf{r}_1-\mathbf{r}_2|} 
    = \frac{e^{-2\lambda |u_1v_2-u_2v_1|}}{2|u_1v_2-u_2v_1|}
\end{equation}
And the normalization factor:
\begin{equation}
    N = \langle L,M|L,M \rangle = \int d\Omega_1 \int d\Omega_2\ |u_1|^{2L} |u_2|^{2L} |u_1v_2-u_2v_1|^{4S-2L}
\end{equation}
Pseudopotential $V_{J}$ for any given interaction can be computed as $V_m = \langle L,M| V |L,M \rangle /N $

Make a unitary transformation: $u_2 = u_1 u_2' - v_1^* v_2',\ v_2= v_1 u_2' + u_1^* v_2'$, then the distance is simplified to be:
\begin{equation}
    u_1 v_2 - u_2 v_1 = v_2' (|u_1|^2 + |v_1|^2) = v_2' \notag
\end{equation}
This induces an orthogonal transformation $\mathbf{\Omega} \rightarrow \mathbf{\Omega}'$ on the points of the sphere and we have the normalization factor:
\begin{align}
     N = \int d\Omega_1 \int d\Omega_2'\ |u_1|^{2L} |u_1 u_2' - v_1^* v_2'|^{2L} |v_2'|^{4S-2L} = \frac{16\pi^2}{(2L+1)! (2S+1)!} \sum_{k=0}^{L} \binom{L}{k}^2
    (L-k)! (L+k)! k! (2S-k)!
\end{align}
The decomposition of Yukawa potential is
\begin{align}
    \frac{1}{2} \langle L,M| \frac{e^{-2\lambda |u_1v_2-u_2v_1|}}{|u_1v_2-u_2v_1|}| L,M \rangle
    &= 32\pi^2 \sum_{k=0}^{L} \binom{L}{k}^2
    \frac{(L-k)! (L+k)!}{2(2L+1)!} 
    \int_0^1 dx\cdot x^{4S-2k} (1-x^2)^{k} e^{-2\lambda x}
\end{align}

Meanwhile, the PP decomposition of e-e interaction is simpler on disk geometry, where is can be decomposed by Laguerre polynomials. In the thermodynamic limit, this formulation becomes equivalent to that in spherical geometry. The two geometries can be related via a stereographic projection.

\section{All-size data of ED for the Laughlin phase with two quasiholes}
To further support the universality of the observed bound state discussed in the main context, we provide a comprehensive finite-size dataset up to the largest system size we can approach -- ranging from 8 to 12 electrons -- under different electron–electron (e-e) interactions. These results show consistent trends across system sizes, further supporting the universality and robustness of the formation of bound states with short-range interactions in Laughlin's case.
\begin{figure}
    \centering
    \includegraphics[width=1\linewidth]{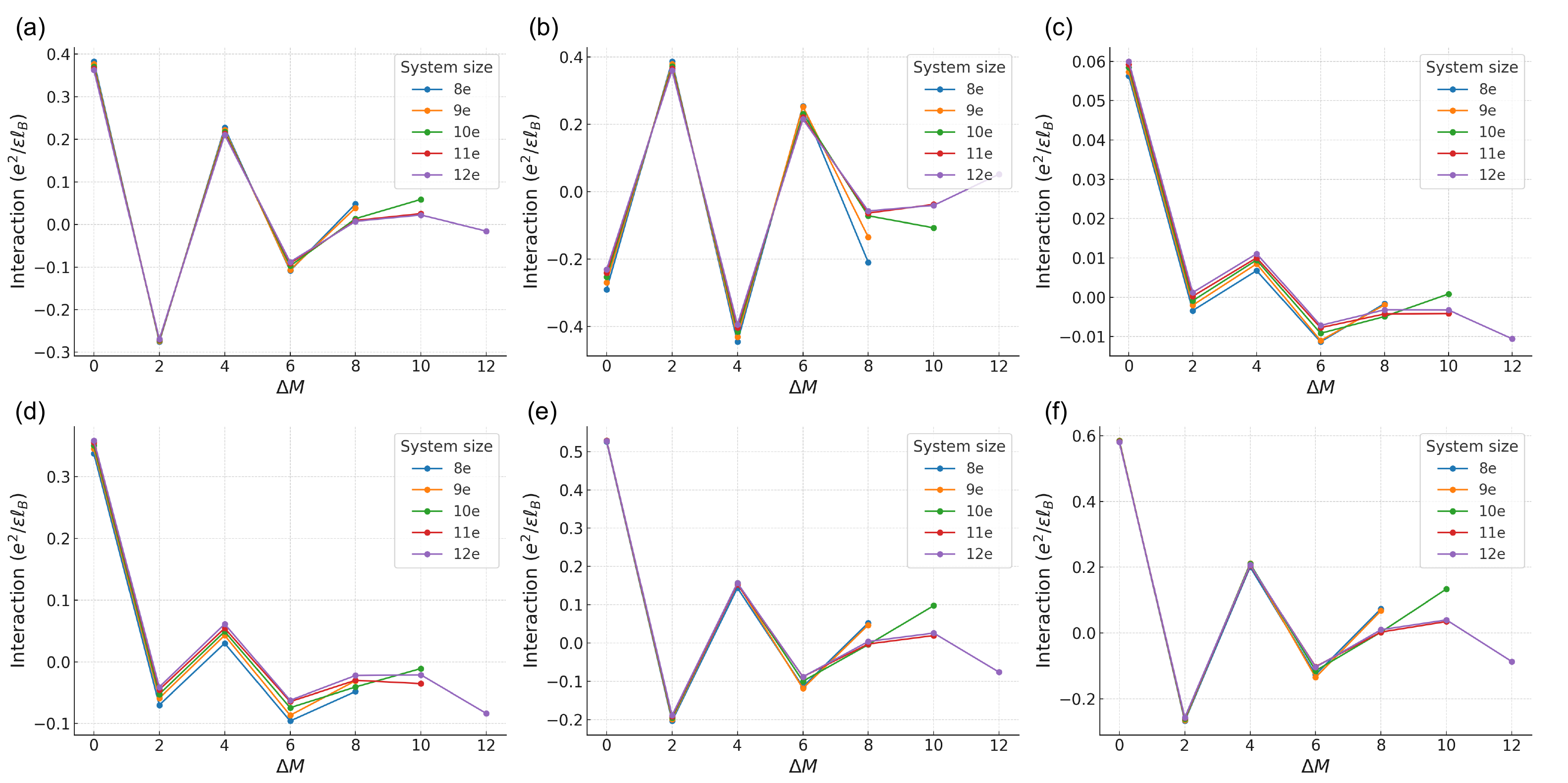}
    \caption{Finite-size evolution of the Laughlin phase's energy with two anyons as a function of their relative angular momentum, for system sizes ranging from $N_e = 8$ to $N_e = 12$ under various e-e interactions: (a) $\hat{V}_3$ and (b) $\hat{V}_5$ in $\mathcal{H_L}$; (c) realistic Coulomb interaction; (d-f) realistic Yukawa interaction with $\lambda = 0.25$, $0.375$, and $0.5 l_B$, respectively.}
    \label{fig:placeholder}
\end{figure}

\section{Comparison of the Energy Scale with Shot Noise Experiments}
Previous shot noise experiments in the Laughlin 1/3 systems show a temperature-dependent crossover in the effective charge, changing continuously from $e^* = 2e/3$ at the lowest achievable temperature $\sim 10mK$ to $e/3$ when elevating the temperature.  We will provide a scale argument based on our numerical data and the two-anyon “molecule” model explained in the main context.

In the computation of the effective interaction between two Laughlin 1/3 anyons induced by electron-electron Yukawa interaction when $\lambda = 0.375/l_B$, we set its coefficient of $\hat{V}_1$ the same as its value in Coulomb and rescale other coefficients accordingly to compare the different impact of these two interactions more directly. But here to relate to the real experiment, the initial values of the coefficients should be retrieved.

So the effective energy with Yukawa interaction shown in Figure 1 (b) of the main text is enlarged around 100 times. Its real value in the system where the interaction is realized should be $\sim 10^{-4} e^2/\epsilon l_B$. Taking the dielectric constant $\epsilon = 3\sim 5 \epsilon_0$, $\epsilon_0$ is the dielectric constant in vacuum, and magnetic filed $B\sim 10 T$, the bounding energy for the $\Delta M =2$ anyon “molecule” is $E_{\Delta M =2} \approx 10^{-25}\sim 10^{-24} J$. 

The thermal energy at temperature $T$ is given by $k_B T$, where $k_B$ is the Boltzmann constant. At $T \sim 100mK$:
\begin{equation}
    k_B T = 1.38 \times 10^{-23} \text{J/K} \times 0.1 \text{K} = 1.38 \times 10^{-24} \text{J}
\end{equation}
This is comparable to the estimated binding energy of the two-anyon molecule in the realistic LLL Yukawa interaction regime. Hence, at temperatures around $100 \text{mK}$, thermal fluctuations are expected to compete with and overcome the bound-state energy, leading to thermal dissociation of anyon molecules and the recovery of quasiparticle tunneling with effective charge $e/3$. At lower temperatures (e.g., $T\sim 10 \text{mK}$ when $k_B T \sim 1.38 \times 10^{-25} \text{J}$), the binding energy dominates over thermal fluctuations, making the formation of two-anyon clusters with effective charge $2e/3$ thermodynamically favourable. This provides a natural explanation for the temperature-dependent crossover of effective charge observed in shot noise experiments.

\section{Energy spectrum for three-anyon clusters in Laughlin 1/3}
\begin{figure*}[h]
    \centering
    \includegraphics[width=0.9\linewidth]{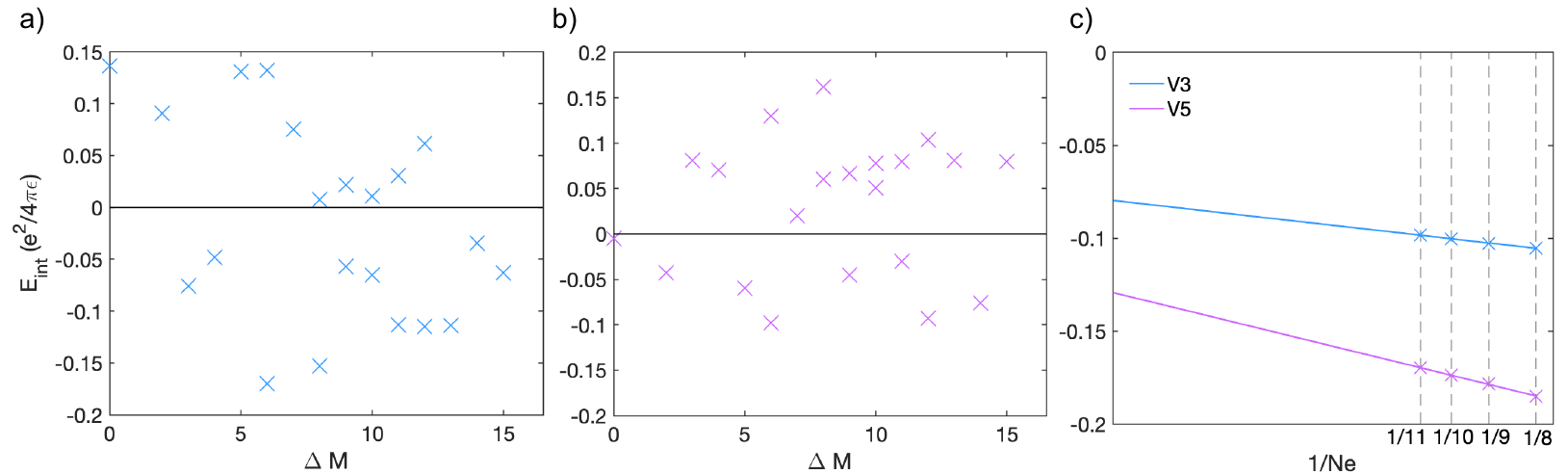}
    \caption{Effective interaction of 3-anyon cluster under model Hamiltonian (a) $\hat{V}_3$ and (b) $\hat{V}_5$ in $\mathcal{H_L}$. (c) Finite-size scaling of the interaction energy with model Hamiltonian $\hat{V}_3$ (blue) and $\hat{V}_5$ (purple) at $\Delta M = 6$ in a cluster of three Laughlin 1/3 anyons.}
    \label{fig_l3qh_model}
\end{figure*}

Figure \ref{fig_l3qh_model} (a) and (b) show the effective interacting energy, whose definition has been explained in the main text, within a three-anyon cluster in Laughlin 1/3 FQH phase introduced by model Hamiltonian $\hat{V}_{int} = t \hat{V}_1 + \hat{V}_{2n+1}, \ n = 1, 2$ ($\hat{V}_{3}$ and $\hat{V}_{5}$). $t\gg 1$ is required to make sure that we are inside the $\mathcal{H_L}$. Here the number of quasiholes $N_{qh} = 3$. For a multi anyon system ($N_{qh} \geq 3$), there are two-body interactions in-between each pair of anyons while the CHS also allows the existence of effective many-body interactions. So the constituents of the effective interaction in a three-anyon cluster are three pieces of two-body interactions and the three-body interaction among the whole cluster. We will discuss the entire effective interaction as a whole and not dig into details to separate and clarify the effects from its different components in this work. 

In both cases for $\hat{V}_{3}$ and $\hat{V}_{5}$, the lowest energy state is $\Delta M = 6$ state where the effective interactions are negative, representing an energetically favored $\Delta M = 6$ three-anyon cluster. The finite-size scaling of the value for this effective interacting energy at $\Delta M = 6$ in Figure \ref{fig_l3qh_model} (c) verifies this negativity at thermal dynamic limit.

\section{Self Energy of MR $e/4$ quasiholes}

\begin{figure*}
    \centering
    \includegraphics[width=1\linewidth]{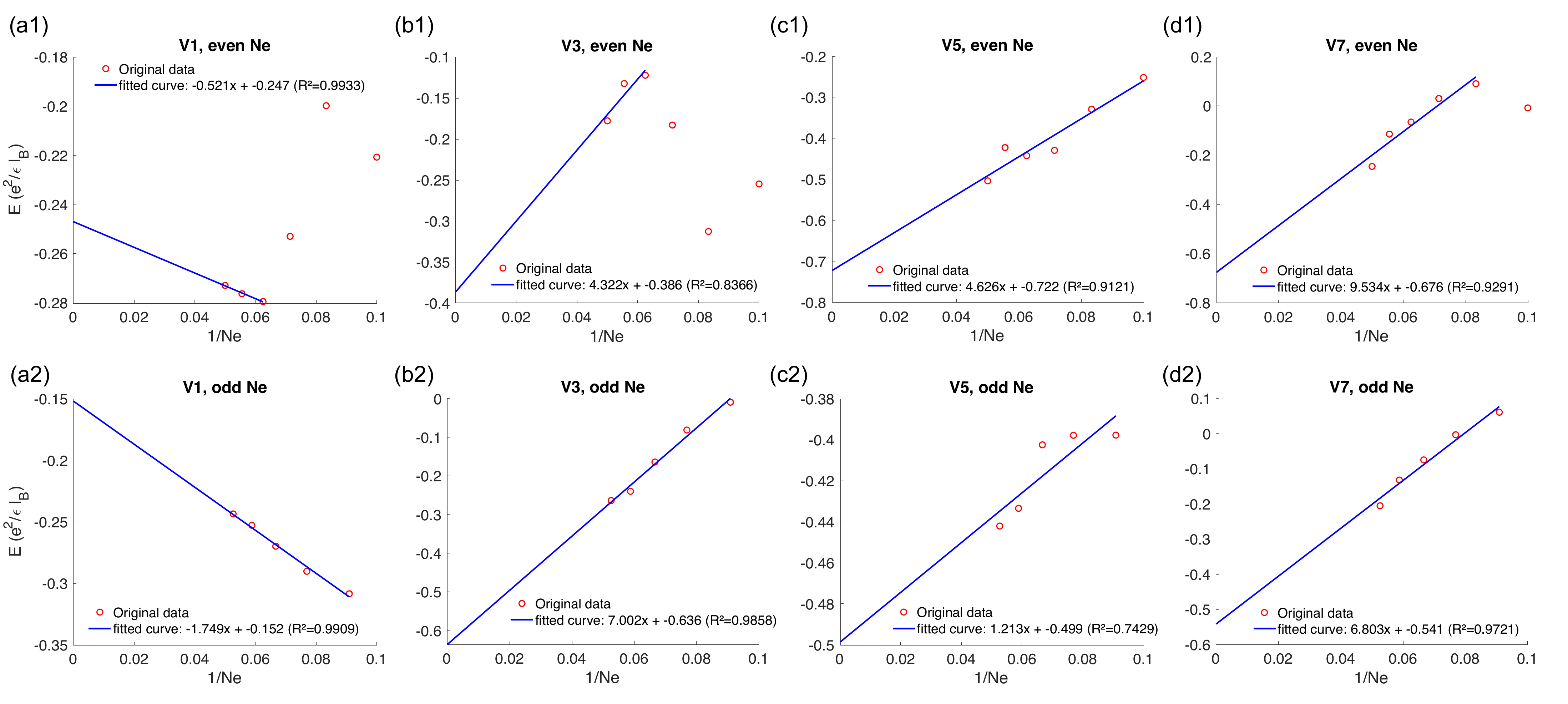}
    \caption{Finite-size scaling of the self energies of the MR quasiholes induced by PPs.}
    \label{fig_mr_selfE}
\end{figure*}

Inserting a single flux quantum into the MR ground state creates two $e/4$ MR quasiholes, a consequence of flux fractionalization in non-Abelian phases. The total energy of the system then have three contributions: the ground state energy $E_g$, twice the self-energy of a single $e/4$ MR quasihole, and the interaction energy between the two anyons. Unlike the Laughlin case, the self-energy cannot be extracted directly due to the coexistence of interactions between the anyons. However, by placing the two quasiholes at the North and South poles of the sphere, their interaction energy decays to zero as the system size increases and the separation becomes sufficiently large.

The effective energy associated with the quasiholes and their interaction is defined as
\begin{equation}
    E_{\text{eff}} =  N_{qh}E_{\text{qh}} + E_{\text{int}} = E_{\text{total}} - E_\text{g}, \quad N_{qh}=2
\end{equation}
where $E_{\text{total}}$ is the total energy of the system, $E_{qh}$ is the self-energy of a single $e/4$ quasihole, and $E_{\text{int}}$ denotes the interaction energy between the quasiholes. The self-energy of a single quasihole is then approximated by extrapolating the effective energy in the thermodynamic limit: 
\begin{equation}
    E_{\text{qh}} = \lim_{N_e \to \infty} \frac{1}{2} E_{\text{eff}}
    \label{eq_mr_selfE}
\end{equation}
It is assumed that the self-energies of the two anyons in a system with a fixed topological number (“$1$” or “$\psi$”) are identical. However, since the self-energy may differ between even and odd $N_e$ systems, the two cases are treated separately.

Figure \ref{fig_mr_selfE} shows $E_{\text{eff}}/2$ plotted as a function of $1/N_e$ for odd and even $N_e$ systems with various model electron-electron interactions. The data points are extrapolated to extract the self-energies according to Eq.(\ref{eq_mr_selfE}), using the largest system sizes accessible within our numerical limits--$N_e = 19$ for odd-parity and $N_e = 20$ for even-parity systems. We note that the extraction of self-energies is strongly constrained by the system-size limitations of exact diagonalization. Since larger systems yield weaker interactions and thus more reliable data, the fitting and extrapolation are weighted more heavily toward results from larger systems. 

Additionally, the behaviours of $E_{\text{eff}}$, and consequently the self-energies of the $e/4$ MR quasiholes, exhibit a more stable and systematic trend in odd-$N_e$ systems (topological quantum number “$\psi$”) compared to even-$N_e$ systems (topological quantum number “$1$”). This also implies that the neutral Majorana fermion mode exhibits a more stable self-energy than the charged 
U(1) bosonic mode according to the bulk-dedge correspondance. The electrically neutral Majorana mode is less sensitive to interactions and finite-size effects \cite{MooreRead1991, ReadGreen2000}, whereas the charged U(1) sector couples strongly to background potentials and orbital discretization \cite{RezayiHaldane2000}. Moreover, the Majorana mode is topologically protected as part of the non-Abelian order, reducing its susceptibility to perturbations \cite{MooreRead1991, Wen2004}. Similar conclusions regarding the robustness of neutral edge modes have been discussed in studies of edge structures \cite{MilovanovicRead1996}.

\section{Energy spectrum for four-anyon clusters in Moore-Read phase}

\begin{figure*}[h]
    \centering
    \includegraphics[width=1\linewidth]{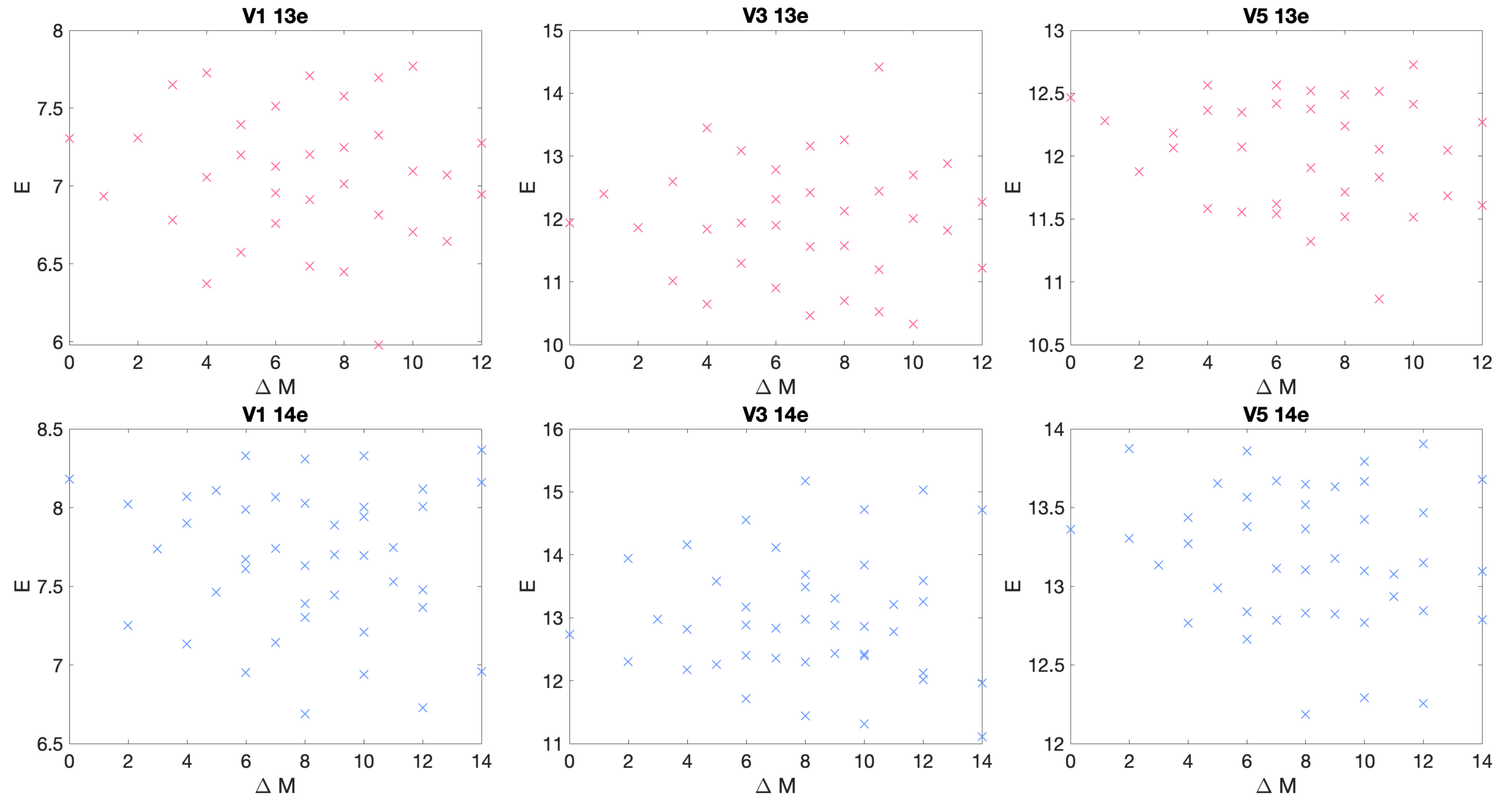}
    \caption{Energy spectrum of 4-anyon cluster under model Hamiltonian $\hat{V}_1$, $\hat{V}_3$, and $\hat{V}_5$ in both odd and even $N_e$ systems}
    \label{fig_mr4qh_model}
\end{figure*}

Figure \ref{fig_mr4qh_model} shows the energy spectrum of quasihole excitations for systems with odd ($N_e = 13$, top row) and even ($N_e = 14$, bottom row) parities with electron-electron model interaction $\hat{V}_1$, $\hat{V}_3$, and $\hat{V}_5$ inside the MR CHS. Across all cases, the lowest-energy states are located at relatively large $\Delta M$. The counting, which is one of the reflection of the topological property of the system, is distinct in between systems hold the two different parities\cite{boyang-2021-statistical}. 

As a transition from model Hamiltonian to realistic ones, we studied the low-lying energy spectrum of MR four-quasihole systems with the model Hamiltonian Eq.(\ref{eq_mr_changeT}) in 1LL.
\begin{equation}
    \hat{H}_{int} = t \hat{V}_3^{3bdy} + \hat{V}_{\text{Coulomb}}
    \label{eq_mr_changeT}
\end{equation}
where $\hat{V}_3^{3bdy}$ is the three-body pseudopotential that defines the MR conformal Hilbert space $\mathcal{H_{MR}}$. When $t\to \infty$, the system is entirely within the $\mathcal{H_{MR}}$. The Hilbert space will gradually evolve to the real space as $t$ is tuned down. 

\begin{figure*}[h]
    \centering
    \includegraphics[width=0.9\linewidth]{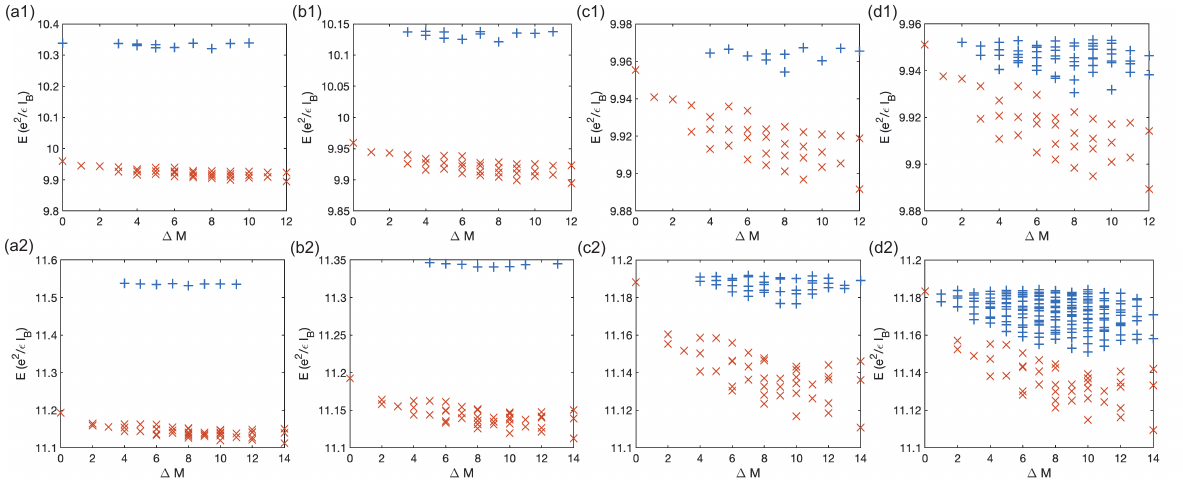}
    \caption{Energy spectrum of the model Hamiltonian.  Red crosses denote the quasihole excitation states while the blue symbols correspond to higher-energy states outside the $\mathcal{H_{MR}}$. First row: odd, second row: even. t decreases from left to right: (a) t =1, (a) t =0.5, (a) t =0.1, (a) t =0.05.}
    \label{fig_mr_changeT}
\end{figure*}

Figure \ref{fig_mr_changeT} shows the evolution of the four-quasihole excitation spectra as $t$ decreases, for both odd (top row) and even (bottom row) $N_e$ systems. Remarkably, the structure of the quasihole spectrum remains largely unchanged even at small $t$, despite the dressing from neutral excitations outside the $\mathcal{H_{MR}}$, as indicated by the little variation in the distribution of the quasihole excitation states and their dependence on the relative angular momentum $\Delta M$. 

\begin{figure}[h]
    \centering
    \includegraphics[width=0.5\linewidth]{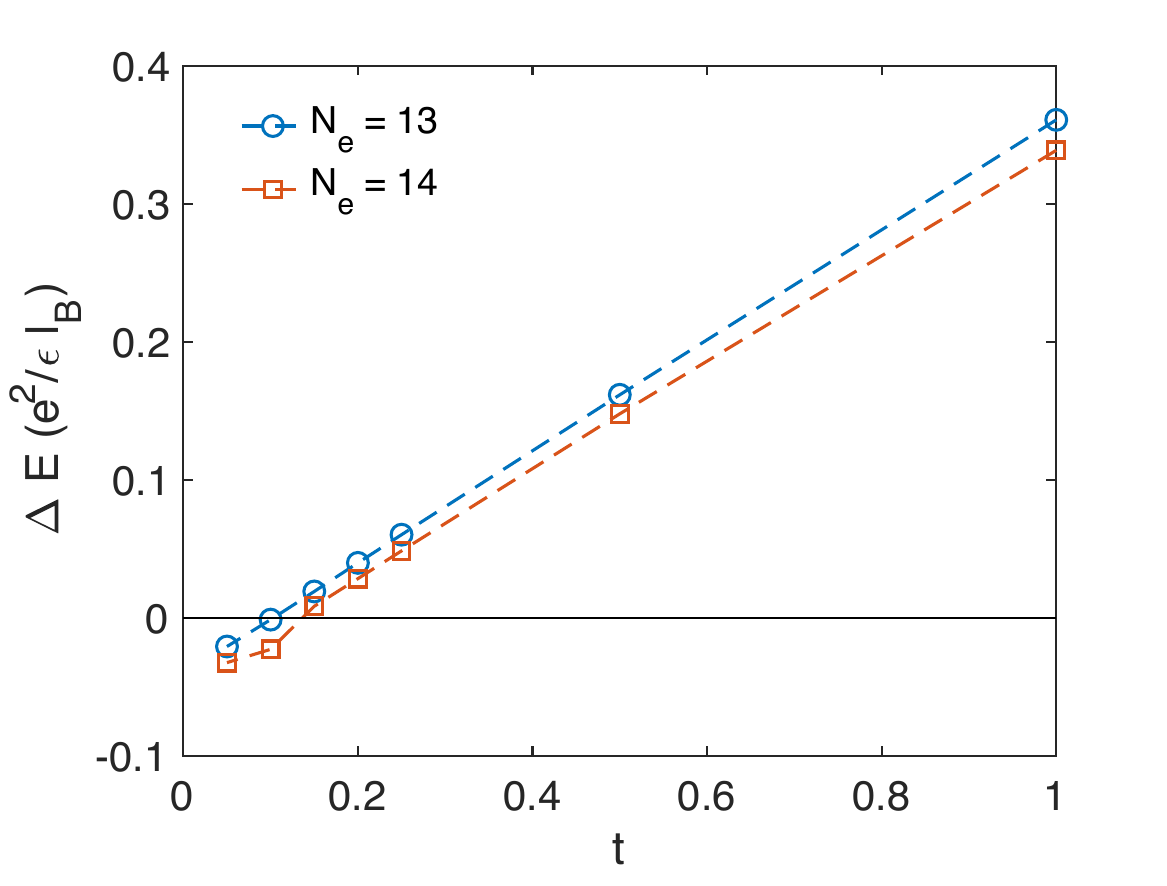}
    \caption{Energy gap between quasihole excitations and other parts in the spectrum of the four-quasihole MR phase with model Hamiltonian \ref{eq_mr_changeT} plotted against t.}
    \label{fig_mr_change_t_gap}
\end{figure}

The energy gap $\Delta E$ between the quasihole excitation manifold and the continuum as a function of $t$ is shown in Figure \ref{fig_mr_change_t_gap}. For both $N_e = 13$ and $N_e = 14$ systems, the gap exhibits a almost perfectly linear dependence on $t$ and remains open until $t$ is reduced to a very small value. This indicates that the quasihole structure within the MR conformal Hilbert space is largely preserved across a wide range of $t$. And only when $t$ becomes sufficiently small, where the system effectively approaches the full realistic Hilbert space limit, does the topological gap finally collapse, signaling the breakdown of the gapped phase.

\section{Spatial extent analysis}

The estimated radius of a few-anyon cluster $R_1$ (in units of $l_B$) from the first moment is given by:
\begin{equation}
    R_1 = \sqrt{S}\ \frac{\int_0^\pi |\delta \rho(\theta) | \theta \sin{\theta} d\theta}{\int_0^\pi |\delta \rho (\theta) | \sin{\theta} d\theta}
    \label{eq_1st_moment}
\end{equation}
where $\delta \rho(\theta) = \rho(\theta)- \rho_0$, and $\rho_0$ is a background density. We take $\rho_0$ to be the ground-state density without additional fluxes in our analysis. The second moment $R_2$ in units of $l_B$ is:
\begin{equation}
    R_2 = \sqrt{S} \sqrt{ \frac{\int_0^\pi |\delta \rho| \theta^2 \sin{\theta} d\theta}{\int_0^\pi |\delta \rho| \sin{\theta} d\theta} }
    \label{eq_2nd_moment}
\end{equation}
Eq. (\ref{eq_1st_moment}) and (\ref{eq_2nd_moment}) give two model-independent estimate of the arc lengths for a few-anyon cluster on sphere. The arc length will asymptotically approach the radius of the anyon cluster on planar geometry in the thermodynamic limit.

\subsubsection{Laughlin}
The two radii for the $\Delta M = 12$ three-anyon cluster with electron-electron Coulomb interaction, computed in $N_e = 11$ system, are:
\begin{equation}
    R_1^{l} \approx 6.28 l_B,\ R_2^{l} \approx 6.92 l_B
\end{equation}
Typically, the magnetic length is estimated as $l_B\sim 26\text{nm}/\sqrt{B}$, where $B$ is the strength of the magnetic field in Tesla. The Laughlin 1/3 state is realized at magnetic fields of $B\sim10-15 \text{T}$ in experiments. These values provide an estimate for the spatial extent of the anyon cluster as: $R_1^{l} \sim 40-50 \text{nm}$, $R_2^{l} \sim 45-60\text{nm}$. Similarly, for the Yukawa interaction with $\lambda = 1$, the lowest-energy $\Delta M = 6$ state has the first and second moments: 
\begin{equation}
    R^{l'}_1 \approx 6.38 l_B,\ R^{l'}_2 \approx 7.23 l_B
\end{equation}
which correspond to the spatial extents: $R^{l'}_1 \sim 40-50\text{nm}$, and $R^{l'}_2 \sim 45-60\text{nm}$. 

The calculated spatial extents of anyon clusters from the first and second moments are unexpectedly similar under both Coulomb and Yukawa interactions. This similarity arises because the moment-based calculation primarily captures the weighted average separation of the density, where both the core region and oscillatory tails contribute significantly. Although the density profile under Yukawa interaction exhibits core dominance, while the Coulomb profile features a more oscillatory structure, the global contribution from the entire density profile ultimately yields comparable spatial extents (for density profiles and more detailed discussions, see the Supplementary). Thus, the spatial extent calculation reflects the global density distribution, while $\Delta M$ more accurately captures the cluster compactness and inside structures.

\begin{figure}
    \centering
    \includegraphics[width=0.8\textwidth]{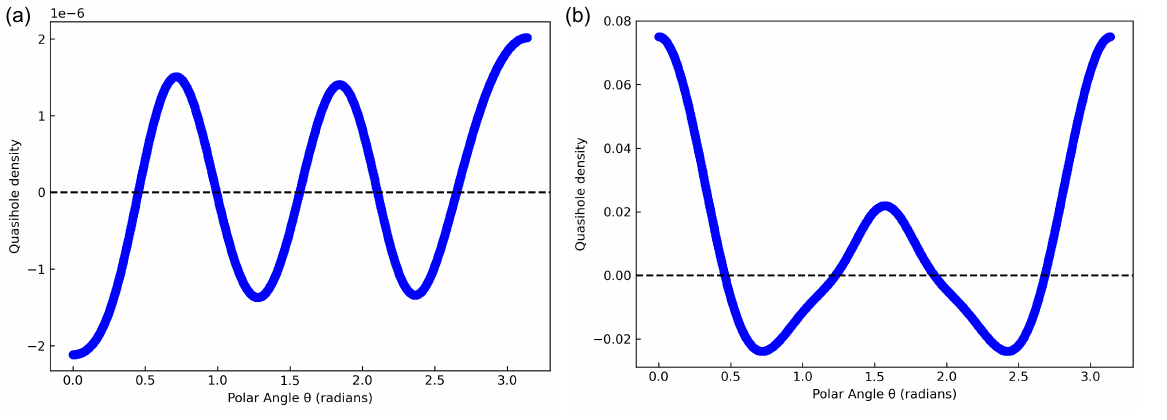}
    \caption{Quasihole density for (a) the $\Delta M = 12$ cluster with Coulomb interaction, and (b) Yukawa interaction ($\lambda = 1$). $N_e = 12, N_o=36$.}
    \label{fig_l_density}
\end{figure}

The density profile of three-anyon clusters under electron-electron Coulomb interaction in Laughlin phase exhibits multiple sharp peaks, indicating long-range correlations and oscillatory structures (Fig \ref{fig_l_density} (a)). This pattern reflects weak central binding, with the cumulative contribution rising gradually and reaching the 50\% mark  at a large polar angle ($\theta \approx 2.0$ radians), suggesting that the density deviation is more evenly spread over the sphere without a dominant core. In contrast, the short-range Yukawa interaction produces a density profile with a prominent central peak and mild, smooth oscillations extending toward the edges (Fig \ref{fig_l_density} (b)). This indicates that the central region holds a significant fraction of the density deviation, reflecting strong short-range binding. The cumulative contribution shows that the core region (within $\theta \approx 1.5$ radians) already accounts for about 50\% of the total density deviation. While the core is compact, the long, smoothly decaying tail contributes notably, leading to a larger-than-expected radius.

The moment-based calculation reflects the global density distribution, giving substantial weight to both the central peak and tail contributions. This approach does not fully capture the difference in cluster compactness or the detailed structures arising from different electron-electron interactions. Although the Yukawa interaction forms a tightly bound core (as indicated by both the density profile and the reduced relative angular momentum $\Delta M$), the long tails result in a spatial extent comparable to that of the Coulomb case.

In summary, based on the finite-system study we perform here, the Coulomb density profile is more oscillatory and spread out, while the Yukawa profile shows core dominance with smooth tails. The cumulative contribution plots further demonstrate the differences in spatial distribution, emphasizing the need to distinguish between compactness and global spatial extent when analyzing anyon cluster sizes. Although the calculated spatial radii may appear similar, the underlying clustering mechanisms differ significantly with the two different electron-electron interactions. The spatial extent calculation primarily reflects the averaged density distribution, while $\Delta M$ will capture the internal correlation and cluster compactness.

\subsubsection{Moore-Read}
\begin{figure}[h]
    \centering
    \includegraphics[width=0.8\linewidth]{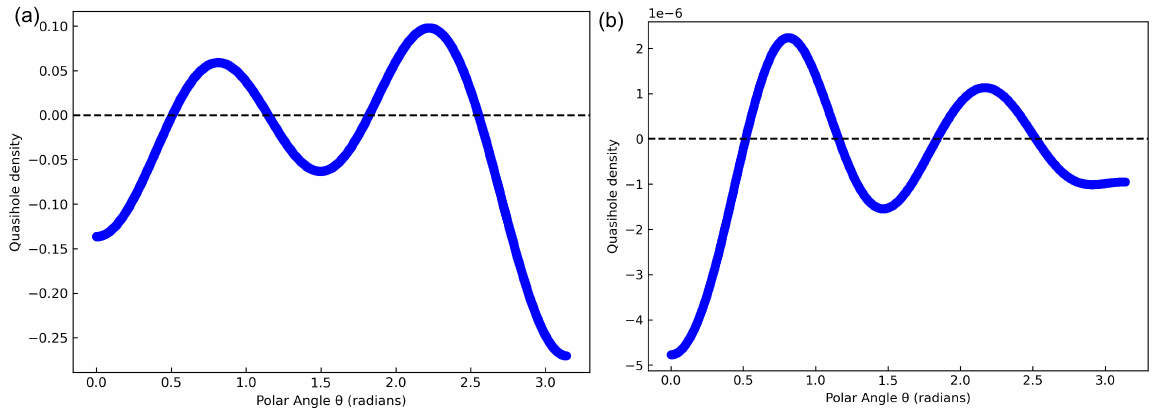}
    \caption{Quasihole density under 1LL Coulomb interaction for (a) $N_e = 15$ and (b) $N_e = 16$ systems, both containing four MR quasiholes.}
    \label{fig_mr_density}
\end{figure}

The quasihole density of MR system containing a four-quasihole cluster in odd and even $N_e$ ($N_e = 15,16$ respectively) systems are shown in Figure \ref{fig_mr_density}. They exhibit similar oscillatory structures with multiple sharp peaks, indicating the long-range structure of the four-anyon cluster with 1LL Coulomb interaction. The radii of the loose anyon cluster are estimated by Eq. (\ref{eq_1st_moment}) and (\ref{eq_2nd_moment}), similarly as in Laughlin's case. For the lowest energy state as $\Delta M = 12$ in $N_e = 15$ (odd) system, the two radii are:
\begin{equation}
    R_1^{MR} \approx 6.77 l_B,\ R_2^{MR} \approx 7.40 l_B
\end{equation}
And for the $\Delta M = 14$ anyon cluster in $N_e = 16$ (even) system,
\begin{equation}
    R_1^{MR'} \approx 5.23 l_B,\ R_2^{MR'} \approx 5.92 l_B
\end{equation}
Experimentally, the $\nu=5/2$ state (partial filling $\nu^*=1/2$ in the 1LL) is realized when the magnetic field $B \sim 2- 3T$. These radii correspond to the spatial extents: $R_1^{MR} \sim 100-125 \text{nm}$, $R_2^{MR} \sim 110-140\text{nm}$; $R_1^{MR'} \sim 75-95 \text{nm}$, $R_2^{MR'} \sim 85-110\text{nm}$. The spatial extent of the anyon clusters with the lowest energy in the MR phase exhibits a difference between odd and even electron systems, suggesting a parity effect in the cluster configuration in the finite systems. And the predicted spatial extents are much larger compared to the Laughlin's cases.

\end{document}